\documentclass[letterpaper,12pt]{article}

\pdfoutput=1
\usepackage{graphicx}
\usepackage{amsmath}
\usepackage{amsfonts}
\usepackage{hyperref}
\usepackage{multirow}

\makeatletter
\renewcommand\section{\@startsection {section}{1}{\z@}%
{-3.5ex \@plus -1ex \@minus -0.2ex}%
{2.3ex \@plus 0.2ex}%
{\normalfont\normalsize\bfseries}}

\renewcommand\subsection{\@startsection{subsection}{2}{\z@}%
{-3.25ex \@plus -1ex \@minus -0.2ex}%
{1.5ex \@plus 0.2ex}%
{\normalfont\normalsize\bfseries}}

\def\@seccntformat#1{\csname the#1\endcsname.\hspace{0.8em}}
\makeatother


\newcounter{example}
\newcommand{\example}[1]{\refstepcounter{example}
\noindent \textbf{Example \theexample: #1}}

\newcounter{assumption}
\newcommand{\assumption}[1]{\refstepcounter{assumption}
\noindent \textbf{Assumption \theassumption: #1}}

\newcounter{context}
\newcommand{\context}[1]{\refstepcounter{context}
\noindent \textbf{Context \thecontext: #1}}

\begin{document}

\setlength{\baselineskip}{3.75ex}
\setlength{\abovedisplayskip}{12pt}
\setlength{\abovedisplayshortskip}{0pt}
\setlength{\belowdisplayskip}{12pt}
\setlength{\belowdisplayshortskip}{12pt}

\noindent
\textbf{\LARGE Analogy making as the basis of statistical}\\[2ex]
\textbf{\LARGE inference}\\[3ex]

\noindent
\textbf{Russell J. Bowater}\\
\textit{Independent researcher,
Doblado 110, Col.\ Centro, City of Oaxaca, C.P.\ 68000, Mexico.\\
Email address: as given on arXiv.org. Twitter/X profile:
\href{https://x.com/naked_statist}{@naked\_statist}\\ Personal website:
\href{https://sites.google.com/site/bowaterfospage}{sites.google.com/site/bowaterfospage}}\\[2ex]

\noindent
\textbf{\small Abstract:}
{\small Standard statistical theory has arguably proved to be unsuitable as a basis for
constructing a satisfactory completely general framework for performing statistical inference.
For example, frequentist theory has never come close to providing such a general inferential
framework, which is not only attributable to the question surrounding the soundness of this theory,
but also to its focus on attempting to address the problem of how to perform statistical inference
only in certain special cases.
Also, theories of inference that are grounded in the idea of deducing sample-based inferences about
populations of interest from a given set of universally acceptable axioms, e.g.\ many theories that
aim to justify Bayesian inference and theories of imprecise probability, suffer from the difficulty
of finding such axioms that are weak enough to be widely acceptable, but strong enough to lead to
methods of inference that can be regarded as being efficient.
These observations justify the need to look for an alternative means by which statistical inference
may be performed, and in particular, to explore the one that is offered by analogy making.
What is presented here goes down this path. To be clear, this is done in a way that does not simply
endorse the common use of analogy making as a supplementary means of understanding how statistical
methods work, but formally develops analogy making as the foundation of a general framework for
performing statistical inference.
In the latter part of the paper, the use of this framework is illustrated by applying some of the
most important analogies contained within it to a relatively simple but arguably still unresolved
problem of statistical inference, which naturally leads to an original way being put forward of
addressing issues that relate to Bartlett's and Lindley's paradoxes.}

\pagebreak

\vspace{4ex}
\section{Introduction}

We may define the fundamental problem of statistical inference (FPSI) as being the problem of
finding a satisfactory completely general framework for performing statistical inference.
Conventional theories of inference, e.g.\ frequentist theory, have never come close to solving this
problem, partially as a result of their focus on attempting to address the problem of how to
perform statistical inference only in certain special cases. Also, although during the second half
of the 20th century, there were high hopes that Bayesian theory would establish itself as being the
widely accepted solution to the FPSI, its inadequacies and limitations have prevented it from
substituting other theories of inference. The aim of the present paper is to clarify how analogies
may be used to perform statistical inference, and by doing so, provide an argument in favour of the
FPSI being solved by means of analogy making.

Assessing the adequacy of a method of inference that is based on analogy making is simply a
question of assessing the adequacy of the analogies that are being made.
This goes against a current trend in favour of assessing the adequacy of a method of inference by
how well it works in practice, e.g.\ by how well calibrated it is when used repeatedly. However, if
this empiricist viewpoint heads in the direction of confidence interval theory, then it will
inherit the defects of that theory.
For example, inferences made about a parameter of interest will only be relevant to a reference
class of cases, which will inevitably include unobserved cases, rather than being directly relevant
to the case at hand, i.e.\ to the data actually observed.
Alternatively, if the empiricist viewpoint being referred to heads in the direction of coverage
checking based on past applications of the method of inference concerned to real, model-simulated
or bootstrapped data, then even assuming that we are able to decide upon a satisfactory overall
measure of how often the uncertainty intervals for all population quantities of interest will
include the true values of these quantities, using this measure to filter out poor methods of
inference is likely to be highly inadequate.
In particular, although some poor methods of inference may be stopped from passing through the
filter in question, what passes through the filter is likely not to only include good methods of
inference, but many poor methods of inference as well.

Among the more serious attempts to solve the FPSI can be found theories of inference that are
grounded in the idea that inferences about a population from an observed sample of the values that
it contains may be deduced from a given set of universally acceptable axioms. Many theories that
attempt to justify Bayesian inference fall into this category, e.g.\ the theories proposed in de
Finetti~(1937), Savage~(1954), Bernardo and Smith~(1994) and Jaynes~(2003), as well as theories
that aim to justify ways of making post-data inferences about population quantities of interest in
terms of imprecise rather than precise probability, e.g.\ the theories proposed in Shafer~(1976)
and Walley~(1991).
However, theories of the general type being referred to that are based on axioms that are strong
enough to justify standard Bayesian inference are open to the criticism that the axioms are too
strong to be universally acceptable, while theories of this type that are based on axioms that are
weak enough to be at least good candidates for being universally acceptable lead to inferences that
could be criticised as being too vague for the method of inference concerned to be regarded as
being efficient.

The preceding discussion therefore justifies looking for a less standard way of trying to resolve
the FPSI, and in particular, looking at the idea of addressing this problem through the use of
analogy making.
While it may be argued that analogy making is used widely as a supplementary means of understanding
how statistical methods work, this paper aims to go further than this and formally develop analogy
making as the foundation of a general framework for performing statistical inference.

Let us now briefly describe the structure of the paper.
In the following section, different types of analogies that may be used for performing statistical
inference will be presented and discussed.
In Section~\ref{sec8}, we will address the issue of the compatibility of analogies of this kind
with contextual information and the compatibility of analogies among themselves when more than one
analogy is being applied to a problem of inference.
Having introduced a reference problem of inference in Section~\ref{sec4}, methods for addressing
this problem based on analogy making will be put forward in Sections~\ref{sec19} to~\ref{sec20},
and the properties of these methods will be discussed in detail.
The final section of the paper (Section~\ref{sec21}), as well as presenting some conclusions about
the use of analogy making as a basis for performing statistical inference, provides further
examples of the clarity that this approach can bring to methodological assessment.

\vspace{3ex}
\section{Types of analogies for performing statistical inference}

\vspace{1.5ex}
\subsection{Fiducial analogy}
\label{sec2}

Before introducing what will be referred to as the fiducial analogy, let us first consider a more
general analogy, which we will call the pre-post event analogy.

\vspace{3ex}
\noindent
\textbf{Pre-post event analogy}

\vspace{1.5ex}
\noindent
This is an analogy between the probability distribution of a random variable before and after the
event of a value being assigned to the variable has taken place. In particular, it is based on the
argument that the probability distribution of a random variable when the event of a value being
assigned to that variable has already taken place should not be judged as being different to what
it was before this event took place if all potential information about what value has been assigned
to the variable is considered to be irrelevant. Therefore, the pre-post event analogy can only be
made if the condition just highlighted, which underlies the argument in question, is valid.
Observe that while the probability distribution of a random variable may be regarded as being a
distribution of physical probabilities before a value has been assigned to this variable, it
arguably must be regarded as being a distribution of subjective or analogical probabilities when a
value has been assigned to this variable, but this value is still unknown. Further discussion of
physical, subjective and analogical probabilities can be found in Bowater~(2022a).

\vspace{3ex}
The pre-post event analogy finds a useful role in statistical inference if the random variable on
which this analogy is based is an appropriately chosen primary random variable (primary r.v.)\
according to the definition of this type of variable given in Bowater~(2019, 2021).
In particular, if nothing or very little was known about the only unknown parameter of the sampling
distribution of a data set of interest before this data set was randomly drawn from the
distribution concerned, then we may well judge that the distribution of an appropriately chosen
primary r.v.\ should be the same, at least for practical purposes, after the data are observed as
it was before the data were observed.
To be clear, the lack of pre-data knowledge about the unknown parameter in question leads the
observed data to be considered irrelevant to judging what distribution should be assigned to the
primary r.v.\ having observed these data.
The pre-post event analogy when applied to a problem of statistical inference in this manner will
be called the fiducial analogy.
With the post-data distribution of the primary r.v.\ having been established, the definition of
this variable may then be used, occasionally with additional assumptions, to obtain the post-data
or fiducial distribution of the parameter of interest.

For the purpose of giving an example of how the fiducial analogy may be used in practice, let us
suppose that we wish to make inferences about the mean $\theta$ of a normal distribution that has a
known variance $\sigma^2$ on the basis of a sample $x$ of size $n$ drawn from the distribution
concerned.
In this example, it is natural to assume that, before the data $x$ are observed, the primary r.v.\
has a standard normal distribution, and to complete the definition of the primary r.v., which we
will denote as $\Gamma$, it is convenient to imagine that the sample mean $\bar{x}$ is determined
before the data set $x$ is generated by substituting the random variable $\Gamma$ with a realised
value of this random variable in the following definition of the distribution of the sample mean in
question:
\begin{equation}
\label{equ1}
\bar{x} = \theta + (\sigma/ \sqrt{n}\hspace{0.1em})\hspace{0.05em}\Gamma
\end{equation}
Given that the sample mean $\bar{x}$ is a sufficient statistic for $\theta$, the process of
generating the data set $x$ conditional on $\bar{x}$ taking the value determined by using this
equation does not depend on the value of $\theta$, and therefore the outcome of this process is
irrelevant from the point of view of making inferences about $\theta$ on the basis of the data $x$.

Under the assumption that nothing or very little was known about the mean $\theta$ before the data
$x$ were observed, we may apply the fiducial analogy to this example, and therefore judge that the
post-data distribution of the primary r.v.\ $\Gamma$ should be the same as the pre-data
distribution of this variable, i.e.\ it should be a standard normal distribution.
By rearranging equation~(\ref{equ1}), it can be seen that making this judgement about the post-data
distribution of the primary r.v.\ $\Gamma$ directly implies that the post-data or fiducial
distribution of $\theta$ is given by the expression:
\begin{equation}
\label{equ20}
\theta\, |\, \sigma^2, x \sim \mbox{N}\hspace{0.05em}(\bar{x},\hspace{0.05em}
\sigma^2\hspace{-0.1em}/n)
\end{equation}

The type of statistical inference that has just been described is referred to as organic fiducial
inference in Bowater~(2019, 2021).

\vspace{3ex}
\subsection{A general sampling analogy}
\label{sec3}

To assign a probability distribution to a variable that has a fixed but unknown value, an analogy
may be made between our uncertainty about this value in the real-world scenario of interest and
what our uncertainty about this value would be if it was about to be randomly drawn from a
population of known composition, i.e.\ the distribution of the values that make up this population
is known.
Using a sampling analogy of this kind to assign a probability distribution to a variable of
interest is the basis of a type of probability called analogical probability, which was alluded to
in the previous section and which is described in detail in Bowater~(2022a).
In statistical inference, this type of analogy may be used either directly or indirectly to perform
the task of determining a post-data distribution for the parameters of the sampling distribution of
a data set of interest, and indeed, it could be argued that in order to perform this task, the
explicit or implicit use of this analogy can not be avoided.
For example, it is clear that, in making inferences about a parameter of interest using what, in
the previous section, was defined to be the fiducial analogy, the type of sampling analogy under
discussion is used indirectly to determine the post-data distribution of the parameter concerned.

Also, we may use the general sampling analogy just described to determine a probability
distribution for the fixed but unknown parameters of a sampling distribution of interest that
represents what we know about these parameters before we try to learn more about them by generating
a data set from the sampling distribution concerned, i.e.\ this analogy may be used to determine a
pre-data or prior distribution for these parameters. In this context, the analogy in question will
be referred to as a pre-data parameter sampling analogy, while in the previous context, this
analogy may be classified as a post-data parameter sampling analogy.

\vspace{3ex}
\subsection{Bayesian analogy}
\label{sec1}

If, as an expression of our pre-data knowledge about the parameters of a sampling distribution of
interest, we decide that the pre-data parameter sampling analogy that was just described should
remain active after a data set has been randomly drawn from this distribution, then the pre-data or
prior distribution for these parameters that results from using this analogy may be used in
combination with Bayes' theorem and the sampling distribution concerned to obtain a post-data or
posterior distribution for these parameters.
Moreover, making the decision to maintain the analogy in question active after observing a given
data set as an expression of our pre-data knowledge about the parameters of interest justifies the
use of Bayes' theorem in the case where prior probabilities are subjective rather than physical
probabilities, and is arguably the only sensible way of justifying the use of Bayes' theorem when
prior probabilities are of this nature.
In this context, we will therefore refer to the pre-data parameter sampling analogy under
discussion as being the Bayesian analogy.
Other ways of trying to justify the use of Bayes' theorem when prior probabilities are subjective
rather than physical probabilities were mentioned in the Introduction and are further discussed in,
for example, Bernardo and Smith~(1994).

By contrast, if we decide that the pre-data parameter sampling analogy just referred to should not
be kept active after observing the data, then as well as still having a justifiable pre-data
distribution of the parameters of interest, we are free to use other analogies to determine a
post-data distribution for these parameters that we may consider as being more appropriate to use
than the Bayesian analogy, e.g.\ the fiducial analogy. Note that the fact that an analogy is no
longer regarded as being active does not imply that the analogy was made incorrectly or
inadequately at a previous time point or in a previous information state.

\vspace{3ex}
\subsection{Other pre-data parameter sampling analogies}
\label{sec6}

In this section, we will consider examples of alternative pre-data parameter sampling analogies to
the one that was considered in the previous section. Similar to the scenario that was just
described in which the sampling analogy being referred to was used to justify the use of Bayes'
theorem when parameters are simply fixed but unknown values, it will be assumed that these
alternative pre-data parameter sampling analogies will be maintained active after observing the
data as expressions of pre-data knowledge about the parameters concerned.

\vspace{3ex}
\example{Moving from the Bayesian to the fiducial analogy}
\label{exa1}

\vspace{1.5ex}
\noindent
Without a great loss of generality, let it be assumed that the sampling distribution from which a
data set of interest was generated depends on only one fixed but unknown parameter. Under this
assumption, an analogy is made between what we knew about the parameter in question before the data
were observed and what would be our uncertainty about the true value of this parameter if it was
about to be randomly drawn from a population with a composition that is either partially or
completely unknown.

\vspace{3ex}
To illustrate the use of this type of analogy, let us imagine that, in a scenario that is
considered to be ideal for the application of the fiducial analogy to a given problem of
statistical inference, we would like to express our uncertainty about the unknown parameter of
interest before the data were observed by using a parameter sampling analogy even though making
this analogy would be surplus to requirement if the fiducial analogy is going to be made. In the
analogy being referred to, it will be assumed, as we have done up until now, that the true value of
the parameter is about to be sampled from a population of possible values in a random manner.
Under this assumption, it should be clear that, in order for the analogy in question to be
consistent with the scenario being ideal for the application of the fiducial analogy then, except
for requiring that the population to be sampled only contains values lying within the natural range
of the parameter concerned, the composition of this population would need to be completely unknown.
If this was not the case, then the act of observing the data could give us a reason to adjust the
distribution of the primary random variable, or in other words, making the pre-post event analogy
on which the fiducial analogy is based could be considered as being unacceptable.

Therefore, a clear difference between Bayesian inference as described in Section~\ref{sec1} and
organic fiducial inference as described in Section~\ref{sec2} can be appreciated, as both forms of
inference could be regarded as being based on an expression of pre-data knowledge about an unknown
parameter of interest through the analogy of random sampling from a population of values for the
parameter, but for the former type of inference, the composition of this population must be known,
while for an ideal application of the latter type of inference, the composition of this population
must be completely unknown.

\vspace{3ex}
\example{Weakening Bayesian prior knowledge}
\label{exa2}

\vspace{1.5ex}
\noindent
The analogy that we will now introduce is a special case of Example~\ref{exa1}. In particular, it
will be assumed that in this previous analogy, the composition is known of a subset of the
population from which the true value of the parameter concerned is to be drawn, while outside of
this subset, the composition of the population will be assumed to be either partially or completely
unknown. Clearly, this analogy satisfies the condition that the composition of the population of
values for the parameter that is to be sampled is not completely known as required by the
definition of Example~\ref{exa1}.

\vspace{3ex}
The use of this analogy offers a way of genuinely weakening the strength of the prior information
about a parameter of interest that enters into an analysis of an observed data set through the
application of Bayes' theorem, as opposed to trying to achieve this goal via the standard but
questionable procedure of simply making the prior distribution of the parameter more diffuse over
the real line. How this analogy may be used to achieve this particular goal will be discussed
further in Section~\ref{sec20}.

\vspace{3ex}
\example{Dividing up the parameter space}
\label{exa3}

\vspace{1.5ex}
\noindent
This example is the same as Example~\ref{exa1} except that we know that the population from which
the true value of the parameter concerned is to be sampled was randomly drawn from one of $k$
possible populations $P_1, P_2, \ldots, P_k$ each of which had a known (physical) probability of
being the selected population, but not one of which has composition that is completely known.

\vspace{3ex}
An example of the use of this type of analogy will be presented in Section~\ref{sec22}.

\pagebreak
\example{Imprecise probabilities}
\label{exa4}

\vspace{1.5ex}
\noindent
This example is the same as Example~\ref{exa3} except that the probabilities of the populations
$P_1, P_2, \ldots, P_k$ being the population from which the true value of the parameter concerned
is to be drawn are completely unknown, while, by contrast, the compositions of all of the
populations $P_1, P_2, \ldots, P_k$ are completely known.

\vspace{3ex}
This analogy can be viewed as the basis of how pre-data knowledge about an unknown parameter of
interest is expressed in what, in the field of Bayesian data analysis, is commonly referred to as a
prior to posterior sensitivity analysis, which can be viewed as the idea that motivated the
development of theories of imprecise probability of the type that are discussed in, for example,
Levi~(1985) and Walley~(1991).

To illustrate the use of this analogy, let us consider the case where the unknown parameter of
interest indicates simply whether a given hypothesis $H$ is true or false. Therefore, the
composition of each of the populations $P_1, P_2, \ldots, P_k$ referred to in the description of
Example~\ref{exa4} will be completely specified by the proportion of members of each of these
populations that indicate that the hypothesis $H$ is true.
In this scenario, let $P_{L}$ and $P_{S}$ be the populations in the set $\{P_1, P_2, \ldots, P_k\}$
having, respectively, the largest proportion $p_L$ and smallest proportion $p_S$ of members that
indicate that the hypothesis $H$ is true.
With these definitions having been made, we may observe that, since the probabilities of the
populations $P_1, P_2, \ldots, P_k$ being the population that is to be sampled are completely
unknown, the analogy being called Example~\ref{exa4} effectively reduces to one where there are
only two possible populations, namely $P_L$ and $P_S$ as just defined, with the probability that
any given one of these two populations is the population to be sampled being completely unknown.

If this analogy is maintained active as an expression of pre-data knowledge about the truth of
hypothesis $H$ after observing a data set drawn from the given sampling distribution of interest,
then it would be natural to make inferences about the truth of hypothesis $H$ by separately
treating each of the proportions $p_L$ and $p_S$ just defined as being the prior probability of the
hypothesis $H$ being true, and using Bayes' theorem to calculate the corresponding posterior
probability of $H$ being true.
The two values for the posterior probability in question that result from doing this could then be
treated as upper and lower limits on the posterior probability of the hypothesis $H$ being true.
Therefore, simply representing our pre-data knowledge about the truth of hypothesis $H$ using the
analogy being discussed naturally leads us to make inferences about the truth of the hypothesis $H$
after the data of interest are observed by performing what is commonly referred to as a prior to
posterior sensitivity analysis.

Clearly, when the unknown parameter of interest is a continuous rather than a binary variable, the
derivation of inferences about the parameter after the data are observed on the basis of the
analogy under discussion will be a little more complex than the simple case just examined.
Nevertheless, for any given function of the parameter concerned, upper and lower limits for the
expected value of this function over the posterior distribution of the parameter may be obtained in
a way that is in accordance with this analogy by using Bayes' theorem as part of a standard prior
to posterior sensitivity analysis.

The usefulness of this type of analogy as part of a method for making inferences about a given
parameter of interest on the basis of an observed data set will depend on how adequately pre-data
knowledge about the parameter may be expressed through the use of this analogy, and how well the
method performs in ensuring that inferences made about the parameter reflect an efficient use of
all the information about the parameter that is contained in the data set of interest.
It can be argued that the method of inference that is being referred to does not score highly in
terms of these two criteria for evaluating its practical merit, and this becomes more apparent when
competing methods of inference are properly taken into account, e.g.\ methods based on the fiducial
analogy or the Bayesian analogy.
Of course, if the posterior distribution of an unknown parameter of interest is insensitive to
which population in the set $\{ P_1, P_2, \ldots, P_k \}$ is regarded as defining the prior
distribution of the parameter, then the analogy under discussion will often be considered as having
substantial merit in the case where the main alternative to using this analogy is viewed as being
the use of the Bayesian analogy.
However, outside of this special case, this type of analogy can arguably be considered to be of
limited practical use as a basis for performing statistical inference.

\vspace{3ex}
\subsection{Artificial data analogies}
\label{sec7}

In a situation where an analogy is already being used for the purpose of making inferences about an
unknown parameter $\theta$ of a sampling distribution of interest on the basis of a data set $x$
drawn from this distribution, e.g.\ the fiducial or Bayesian analogy, we may choose to use an
additional analogy in order to incorporate pre-data knowledge about the parameter $\theta$ into the
inferential process.
In particular, we may choose to make an analogy between the knowledge that we had about the
parameter $\theta$ before the data set $x$ was observed and the information that we would have
obtained about this parameter from observing a preliminary data set $y$ before observing the data
set of main interest $x$. To adequately construct this analogy, we would clearly need to decide
what data values should be included in the preliminary data set $y$ so that the information
contained in this data set about the parameter $\theta$ would best represent the pre-data knowledge
that we had about this parameter. Having made this analogy, a method of statistical inference that
is appropriate for the scenario of interest may then be applied to the data set that results from
combining the artificial preliminary data set $y$ and the data set actually observed $x$ with the
intention that post-data inferences about the parameter $\theta$ are made that not only reflect the
information about the parameter contained in the data set $x$, but also pre-data knowledge about
this parameter.

In the case where the Bayesian analogy is to be used to make inferences about the parameter
$\theta$, the information about this parameter contained in the artificial data set $y$ would be
completely represented by the likelihood function of $\theta$ that is based on having observed the
data set $y$.
If this likelihood function is combined with a preliminary choice for the prior distribution of the
parameter $\theta$ in accordance with Bayes' theorem, then a posterior distribution of this
parameter may be obtained that represents our prior knowledge about the parameter $\theta$ before
the data set of main interest $x$ is observed, and which therefore may be used as a prior
distribution of $\theta$ in a Bayesian analysis of the data set $x$.
Nevertheless, supplementing the Bayesian analogy with an artificial data analogy in the manner
being referred to would seem to be of limited practical value as it would appear to be generally
more convenient to elicit the prior distribution of the parameter $\theta$ for analysing the data
set $x$ directly rather than via the two-step procedure just described.

Let us now assume that a post-data density function for a parameter of interest $\theta$, which we
will denote as the density $f(\theta\,|\,x)$, is determined on the basis of an observed data set
$x$ using no other analogy apart from the fiducial analogy, and also a post-data density function
for this parameter, which we will denote as the density $f(\theta\,|\,x,y)$, is determined on the
basis of the observed data set $x$ combined with an artificial preliminary data set $y$ using again
just the fiducial analogy.
For the analysis of the data set $x$, we may now define what, in Bowater~(2019, 2021), is referred
to as the global pre-data (GPD) function of the parameter $\theta$ as follows:
\vspace{1ex}
\[
\omega_G(\theta) = \left\{
\begin{array}{ll}
a f(\theta\,|\,x,y)/f(\theta\,|\,x)\, & \mbox{if $f(\theta\,|\,x)>0$}\\[1ex]
0 & \mbox{otherwise}
\end{array}
\right.
\vspace{1ex}
\]
where $a$ is any given positive constant.
The GPD function $\omega_G(\theta)$ can therefore be viewed as an adjustment function that weights
the post-data density $f(\theta\,|\,x)$ such that normalising the resulting weighted density to
integrate to one yields the post-data density $f(\theta\,|\,x,y)$.

Under this interpretation, it is clear that, in the context of current interest, the GPD function
$\omega_G(\theta)$ represents all the information that we have about the parameter $\theta$ that is
contained in the artificial data set $y$, and hence, all the pre-data knowledge that we have about
the parameter $\theta$.
Furthermore, in determining the post-data density $f(\theta\,|\,x,y)$, it is clear that we may
bypass the step in which the artificial data set $y$ is directly chosen to represent our pre-data
knowledge about the parameter $\theta$, and instead, take the option of expressing this pre-data
knowledge through the direct elicitation of the GPD function of $\theta$.
Indeed, according to the general definition of a GPD function given in Bowater~(2019, 2021), we may
elicit this function in an unrestricted manner such that it is not obliged to correspond to any
specific artificial data set $y$ that could have been observed. Of course, if we choose to do this,
the post-data density $f(\theta\,|\,x,y)$ would be better denoted as the post-data density
$f(\theta\,|\,x,\mathcal{K}_0)$, where $\mathcal{K}_0$ simply denotes pre-data knowledge that we
had about the parameter $\theta$.

In the more general framework just highlighted, we may interpret the GPD function of a parameter
$\theta$ as being the density function of the parameter $\theta$ determined on the basis of the
data set of main interest $x$ and our pre-data knowledge $\mathcal{K}_0$ about this parameter
relative to (i.e.\ divided by) what this post-data density would be in a reference scenario where
nothing or very little was known about the parameter $\theta$ before the data $x$ were observed.
Finally, we may generalise the framework in question even further and allow the post-data density
of $\theta$ to be derived in this reference scenario using either the fiducial analogy on its own
or in combination with the Bayesian analogy. Further details about what is being referred to here
are given in Sections~2.3 and~3 of Bowater~(2021) and Sections~2.3 and~11 of Bowater~(2022b).

During the course of the present discussion, we have therefore moved away from using an artificial
data analogy and arrived at a point where, in any given scenario of interest, an analogy is made
with a scenario in which nothing or very little was known about the parameter $\theta$ before the
data set $x$ was observed, and on having obtained a post-data density of $\theta$ that is relevant
to this latter scenario, this post-density is reweighted using a GPD function of the parameter
$\theta$ so that it is relevant to the scenario of interest.
Nonetheless, it has been shown that this kind of reasoning is supported by the original idea of
using an artificial data analogy to represent our pre-data knowledge about the parameter $\theta$.

\vspace{3ex}
\subsection{Statistical modelling analogies}

In statistical inference, it is usually the case that the unknown sampling distribution of a data
set of interest needs to be regarded as belonging to an infinite-dimensional space of possible
sampling distributions or populations from which the data could have been drawn, which may be
conveniently described as an open population space with the term closed population space reserved
to describe this space in the special case where it has a finite number of dimensions.
The standard approach to statistical inference when the population space is open can be viewed as
being based on making an analogy between our genuine uncertainty about where the true sampling
distribution of the data lies in this space and what our uncertainty about this sampling
distribution would be if it was assumed that it must belong to a given parametric model of this
space, i.e.\ a given parametric family of distributions.
Having made such an analogy, inferences about population quantities of interest, e.g.\ the
population mean or variance or population quantiles, are then derived on the basis of the observed
data under the parametric modelling assumption in question, or in other words, under the assumption
that the population space may be treated as being a certain type of closed population space rather
than as being an open population space.

To formalise this way of thinking, it may be imagined that each member of a parametric model of an
open population space represents a given set of sampling distributions that belongs to a given
finite, countable or uncountable partition of the population space concerned.
From this viewpoint, the usefulness of applying the modelling analogy just referred to on the basis
of a given parametric model will depend on how similar the sampling distributions are to each other
in each of the subsets of the open population space that constitute the partition of this space in
question, and how well the sampling distributions that are members of the parametric family of
distributions concerned represent each of these subsets of the population space.
If having taken into account these considerations, the application of the modelling analogy under
discussion on the basis of a given parametric model is considered to be prudent, then by expressing
pre-data knowledge about the true sampling distribution though this parametric model in a way that
may be judged as being satisfactory, it should be possible to make trustworthy inferences about
this sampling distribution on the basis of the observed data of interest.

It may be pointed out that, under this modelling philosophy, the observed data can never bring the
adequacy of the parametric model into question as the sole purpose of the model is to represent our
pre-data uncertainty about the true sampling distribution over the open population space of
interest, and it has effectively just been assumed that it serves this purpose in a satisfactory
manner. The fact that, in practice, a data set of interest may often be considered to raise doubts
about the adequacy of the parametric model being used demonstrates that parametric models are often
applied to data without sufficient thought being given to whether they are sophisticated enough for
the modelling analogy that implicitly underlies their use to be sensibly judged as being an
appropriate analogy to make.

Further development of the philosophy to statistical modelling just discussed can be found in
Bowater~(2024).

\vspace{3ex}
\section{The compatibility of analogies}
\label{sec8}

In the following two sections, we will outline two ways in which analogies used for performing
statistical inference need to be compatible.

\vspace{3ex}
\subsection{Compatibility of analogies with contextual information}

An analogy used for performing statistical inference needs to be compatible with the contextual
information that we have about the problem of inference concerned.
For example, under the assumption that the observed data were randomly drawn from a sampling
distribution belonging to a given parametric family of distributions, the analogy needs to be
compatible with the pre-data knowledge that we had about the parameters of this sampling
distribution.
When more than one analogy is being applied to a problem of inference, it needs to be remembered
that analogies are usually imperfect representations of contextual information, and for this
reason, there may be small inconsistencies between the analogies in question.
However, consistencies of this type will usually be acceptable if it can be ensured that all the
analogies that are being applied to the problem of inference concerned are consistent with a stable
assessment of the contextual information that we have about this problem.

The use of multiple analogies to resolve problems of statistical inference forms a vital part of
the core theory of the fiducial-Bayes fusion, which is presented in
Bowater~(2019, 2021, 2022b, 2023).
For example, in determining a joint post-data distribution of the parameters of a sampling
distribution of interest on the basis of the full conditional post-data distributions of these
parameters, it would seem acceptable to construct some of these full conditional distributions
using the fiducial analogy and others using the Bayesian analogy as is discussed in Bowater~(2023).
Also, in forming a post-data distribution for the only unknown parameter of a discrete sampling
distribution such as a binomial or Poisson sampling distribution, it would seem acceptable to
implement the method proposed in Bowater~(2021), and thereby use the fiducial analogy when it is
assumed that the parameter concerned may take any value in its natural parameter space, but use the
Bayesian analogy when the parameter is conditioned to lie in a narrow closed interval within that
space. Finally, in situations where there was a high degree of pre-data belief that a parameter of
interest may be equal to or lie close to a given value, e.g.\ the value indicative of no treatment
effect, it may well be appropriate to combine the fiducial analogy with the Bayesian analogy to
make inferences about the parameter concerned using the method proposed in Bowater~(2022b).
No further comments about this particular method will be made for the moment as it will be
discussed in detail and developed further in Section~\ref{sec22}.

\vspace{3ex}
\subsection{Compatibility of analogies with each other}
\label{sec11}

As clarified in the previous section, if multiple analogies need to be applied to a problem of
inference in order to resolve it adequately, then some degree of inconsistency between these
analogies is allowable.
However, it is of course not acceptable that the inconsistency between these analogies is so great
that they are incapable of existing together, or in other words, that they are incompatible among
themselves.
Therefore, let us now identify an important way in which a set of analogies may be considered to be
incompatible among themselves in their application to a given problem of inference.

\vspace{3ex}
\noindent
\textbf{A way in which analogies can be incompatible with each other}

\vspace{1.5ex}
\noindent
In trying to resolve a given problem of statistical inference, a set of analogies $\mathcal{A}$
will be considered to be incompatible among themselves if making all these analogies is equivalent
to making another set of analogies $\mathcal{B}$ that could be used to address the same problem of
inference, and yet, the post-data inferences that would be made about the parameters of the
sampling distribution of interest by using the set of analogies $\mathcal{B}$ are not equivalent to
the post-data inferences that would be made about these parameters by using the set of analogies
$\mathcal{A}$.

\vspace{3ex}
An example of this type of incompatibility between analogies will be presented in
Section~\ref{sec23}.

\vspace{3ex}
\section{Reference problem of inference}
\label{sec4}

In the following sections, examples of the use of analogies to perform statistical inference will
be presented. All of these examples will relate to the same generally-defined problem of inference
as it is convenient to have the same benchmark for comparison.

In choosing this reference problem of inference, it was borne in mind that it would be particularly
illuminating to examine a problem to which arguably more than one analogy would need to be applied
in order for it to be resolved satisfactorily, and it would be productive if there was the
possibility that by focusing on how this problem may be addressed, new insights would be gained
about how it could be tackled successfully in a variety of different contexts.
For these reasons, the reference problem of inference was chosen to be the one that will now be
described.

Let us begin by supposing that our overall aim is to make inferences about an unknown parameter
$\theta$ on the basis of a data set $x = \{x_i:i=1,2,\ldots,n\}$ that was randomly drawn from a
sampling distribution with density $g(x\,|\,\theta)$ that depends on the true value of $\theta$.
With regard to pre-data knowledge about the parameter $\theta$, the following assumption will be
made.

\vspace{3ex}
\assumption{Scenario of interest}
\label{asm1}

\vspace{1.5ex}
\noindent
It will be assumed that there was a substantial degree of belief before the data $x$ were observed
in the hypothesis that the parameter $\theta$ would lie in an interval $[\theta_L, \theta_U]$ that
is very narrow in the context of our uncertainty about the true value of $\theta$ having observed
the data $x$.

\vspace{3ex}
Given this generally-defined problem of inference, we will make the following assumption with
regard to the overall strategy that will be used to address this problem.

\vspace{3ex}
\assumption{The global analogy}
\label{asm2}

\vspace{1.5ex}
\noindent
For all approaches to the problem of inference just described that will be considered in this
paper, what was known about the parameter $\theta$ before the data $x$ were observed will be
partially expressed by making an analogy between this pre-data knowledge about $\theta$ and what
would be our uncertainty about the true value of this parameter if it was about to be randomly
drawn from a population of values of which a non-zero proportion $\lambda$ lie in the interval
$[\theta_L, \theta_U]$, where $\lambda$ may be any positive value that we choose.
The distribution of the values in this population that lie outside of the interval $[\theta_L,
\theta_U]$ and the distribution of the values that lie inside this interval may be known, be
partially known or be completely unknown.
Moreover, it will be assumed that, similar to the Bayesian analogy, the analogy in question will be
maintained active after observing the data set $x$ as an expression of our pre-data knowledge about
the parameter $\theta$.
Therefore, the analogy being referred to is a pre-data parameter sampling analogy, according to the
terminology of Section~\ref{sec3}, which has the Bayesian analogy as a special case.

\vspace{3ex}
To justify the practical relevance of the problem of inference just introduced, let us present the
following three simple examples of real-world contexts where this problem may arise.

\vspace{3ex}
\context{Studying a treatment effect}
\label{con1}

\vspace{1.5ex}
\noindent
The parameter $\theta$ is the mean clinical effect of a treatment across patients when there is a
reasonable chance that the treatment has no or a negligible effect on the clinical state of any
given patient.

\vspace{3ex}
\context{Studying the difference between treatment effects}
\label{con2}

\vspace{1.5ex}
\noindent
The parameter $\theta$ is the clinical effect difference averaged across patients between two kinds
of drug treatment when, due to the two drug treatments having the same basic mechanism of action on
the body, there is considered to be a reasonable chance that there will be no or a negligible
difference in the effects of these two treatments on the clinical state of any given patient.

\vspace{3ex}
\context{Studying correlations between performance indicators}
\label{con3}

\vspace{1.5ex}
\noindent
The parameter $\theta$ is the correlation between two variables that measure the performance of a
highly complex system such as the human body when, due to the lack of a clear known mechanism that
links the two variables in the system concerned, there is considered to be a reasonable chance that
there is no or a negligible correlation between these two variables.

\vspace{3ex}
Clearly, in all of the three contexts just described, the interval $[\theta_L, \theta_U]$ as
defined in Assumption~\ref{asm1} would naturally be centred around the value zero.

\vspace{3ex}
\section{Example of the use of the Bayesian analogy}
\label{sec19}

\vspace{1.5ex}
\subsection{Methodological description and an illustration}
\label{sec5}

As a first example of the use of an analogy to try to resolve the reference problem of inference
that was outlined in the previous section, let us consider applying the Bayesian analogy as defined
in Section~\ref{sec1} to this problem.
In using this analogy, we need to go far beyond the global analogy that was introduced by
Assumption~\ref{asm2} in terms of the level of detail that is required about the population from
which the true value of the parameter $\theta$ is to be randomly drawn, since we need, of course,
to specify the exact composition of this population rather than simply the proportion of the values
in this population that lie in the interval $[\theta_L, \theta_U]$, i.e.\ the proportion $\lambda$.
Nevertheless, as explained in Section~\ref{sec1}, once this analogy has been made, we are able to
justify a straightforward application of Bayes' theorem in order to obtain a post-data or posterior
distribution for the parameter~$\theta$.

In particular, by proceeding in this way, the posterior probability that the parameter $\theta$
will lie in the interval $[\theta_L, \theta_U]$, which we will now also denote as the interval $A$,
will be given by the following expression:
\begin{equation}
\label{equ21}
P(\theta \in A\,|\,x) =
\frac{\raisebox{1.5ex}{$\lambda \mbox{\footnotesize $\displaystyle
\int$}_{\hspace{-0.4em}-\infty}^{\infty} g(x\,|\,\theta)\pi(\theta\,|\,\theta \in A)
d\theta$}}{\raisebox{-1.3ex}{$\lambda \mbox{\footnotesize $\displaystyle
\int$}_{\hspace{-0.4em}-\infty}^{\infty} g(x\,|\,\theta)\pi(\theta\,|\,\theta \in A) d\theta +
(1-\lambda) \mbox{\footnotesize $\displaystyle \int$}_{\hspace{-0.4em}-\infty}^{\infty}
g(x\,|\,\theta)\pi(\theta\,|\,\theta \notin A) d\theta$}}
\vspace{1ex}
\end{equation}
where $\pi(\theta\,|\,\theta \in A)$ and $\pi(\theta\,|\,\theta \notin A)$ are the prior densities
of $\theta$ when $\theta$ is conditioned to\linebreak lie in the interval $[\theta_L, \theta_U]$
and when $\theta$ is conditioned not to lie in this interval, respectively, and all other notation
is as defined in Section~\ref{sec4}. Of course, the posterior probability $P(\theta \in A\,|\,x)$
may be calculated less directly by using the following expression:
\vspace{0.5ex}
\[
P(\theta \in A\,|\,x) = \int^{\theta_U}_{\theta_L} \hspace{-0.1em}\pi(\theta\,|\,x) d\theta
\]
where $\pi(\theta\,|\,x)$ is the posterior density of $\theta$ over the whole of the real line that
is obtained in the usual Bayesian way by performing the calculations implied by the following
expression:
\begin{equation}
\label{equ2}
\pi(\theta\,|\,x) = \mathtt{C}_0\hspace{0.05em} g(x\,|\,\theta) \pi(\theta)
\end{equation}
where $\mathtt{C}_0$ is a normalising constant and $\pi(\theta)$ is the unconditioned prior density
of $\theta$.

To illustrate the use of the calculations just mentioned in trying to address the problem of
inference under discussion in a real-world scenario, let us begin by making the following
assumptions.
First, let us suppose that $\theta$ is the population mean and the sampling density
$g(x_i\,|\,\theta)$ of each independent data value $x_i$ is a normal density with mean $\theta$ and
known variance $\sigma^2$.
Second, we will assume that the interval $A=[\theta_L, \theta_U]$ is the interval $[-\varepsilon,
\varepsilon]$, where $\varepsilon$ is a small positive constant, which is an assumption that is
consistent with Contexts~\ref{con1} to~\ref{con3} described in Section~\ref{sec4}.
Third, the height of the prior density of $\theta$ over the intervals $(-\infty,-\varepsilon)$ and
$(\varepsilon,\infty)$ will be assumed to be proportional to a normal density function with a mean
of $\theta_0$ and a variance of $\sigma_0^2$. Finally, we will suppose that the prior density of
$\theta$ over the whole of the real line has the following form:
\begin{equation}
\label{equ3}
\pi(\theta) = \mathtt{C}_1 \phi((\theta-\theta_0)/ \sigma_0)(1+\tau h(\theta))
\end{equation}
where $\mathtt{C}_1$ is a normalising constant, $\phi$ is the standard normal density function, the
den\-sity function $h(\theta)$ corresponds to the expression: $\theta \sim
\mbox{Beta}(4,4,-\varepsilon, \varepsilon)$, i.e.\ $h(\theta)$ is a beta density function for
$\theta$ on the interval $[-\varepsilon, \varepsilon]$ with both its shape parameters equal to 4
and the constant $\tau$ is chosen such that the following condition is satisfied:
\vspace{1ex}
\[
\lambda = \int_{-\varepsilon}^{\varepsilon} \pi(\theta) d \theta
\vspace{1ex}
\]
It can be seen therefore that, in general, the prior density of $\theta$ will be a bimodal
con\-tin\-u\-ous density function, with one mode close to or equal to zero and the other mode close
to or equal to $\theta_0$, which seem to be reasonable properties for this prior density to have
given the nature of the problem of inference that is being addressed.

\begin{figure}[t]
\begin{center}
\includegraphics[width=6in]{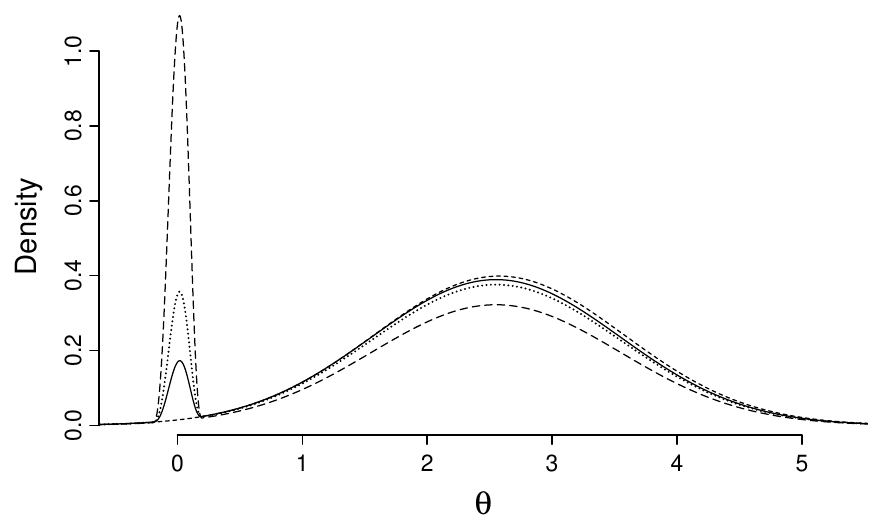}
\caption{\small{Post-data densities of $\theta$ derived assuming that $\varepsilon=0.2$,
$\lambda=0.4$, $\theta_0=0$, $\sigma_0=10$, $\sigma/\sqrt{n}=1$ and $\bar{x}=2.576$ with the
likelihood function of $\theta$ (short-dashed curve) overlaid.}}
\label{fig1}
\end{center}
\vspace{0.5ex}
\end{figure}

\begin{figure}[t]
\begin{center}
\includegraphics[width=6in]{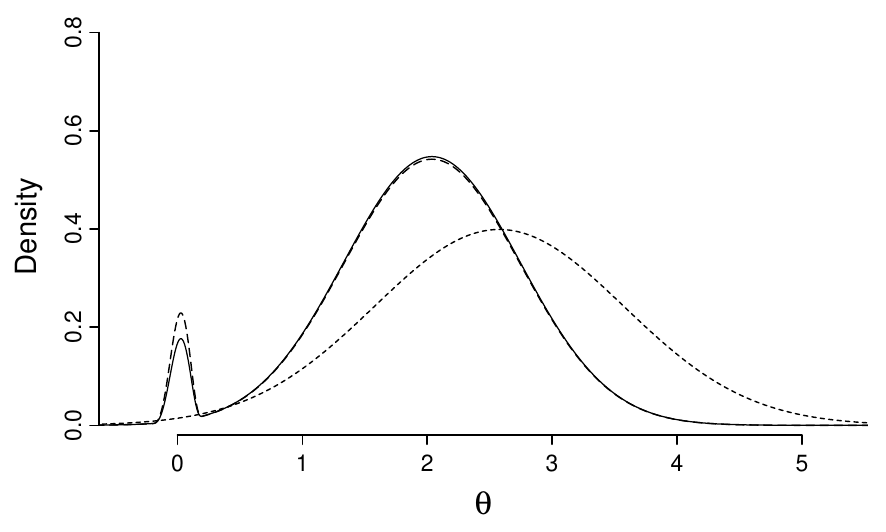}
\caption{\small{Post-data densities of $\theta$ derived assuming that $\varepsilon=0.2$,
$\lambda=0.4$, $\theta_0=1.5$, $\sigma_0=1$, $\sigma/\sqrt{n}=1$ and $\bar{x}=2.576$ with the
likelihood function of $\theta$ (short-dashed curve) overlaid.}}
\label{fig2}
\end{center}
\vspace{0.5ex}
\end{figure}

Under these assumptions, the long-dashed curves in Figures~\ref{fig1} and~\ref{fig2} are plots of
the posterior density of $\theta$ defined by equation~(\ref{equ2}).
In generating both of these plots, it was assumed that the constant $\varepsilon$ is equal to 0.2,
the prior probability that $\theta$ lies in the interval $[-0.2, 0.2]$, i.e.\ the probability
$\lambda$, is equal to 0.4, the observed sample mean $\bar{x}$ is equal to 2.576 and the standard
error of $\bar{x}$ is equal to one.
The assumptions made to derive the two plots being referred to differ due to the fact that, in
Figure~\ref{fig1}, the plot of the posterior density of $\theta$ corresponds to $\theta_0$ being
equal to zero and $\sigma_0$ being equal to 10, while in Figure~\ref{fig2}, the plot of this
density function corresponds to $\theta_0$ being equal to 1.5 and $\sigma_0$ being equal to one.
The short-dashed curves overlaid on Figures~\ref{fig1} and~\ref{fig2} represent the likelihood
function of $\theta$ that corresponds to the assumptions that were just made about the value of the
sample mean $\bar{x}$ and its standard error.

If, in addressing any given problem of inference, it is believed that the Bayesian analogy is an
excellent analogy to make, then of course we should have some degree of confidence in the posterior
distribution of a parameter of interest that is derived on the basis of making this type of
analogy. However, since in trying to resolve the problem of inference under discussion, the
Bayesian analogy will, in practice, usually be far from a perfect analogy to make, two specific
anomalies can easily arise. These two anomalies are usually referred to as Bartlett's paradox and
Lindley's paradox. Let us now describe and discuss both of these paradoxes in detail.

\vspace{3ex}
\subsection{Bartlett's paradox}
\label{sec13}

Bartlett's paradox, as it has become known, is presented in Bartlett (1957), which is a commentary
on Lindley (1957). In this commentary, the same general assumptions are made as in Lindley~(1957).
In particular, as was assumed in the previous section, the sampling density $g(x_i\,|\,\theta)$ of
each independent data value $x_i$ is assumed to be a normal density with mean $\theta$ and known
variance $\sigma^2$.
Also, it is supposed that $\theta_U=\theta_L$ and the prior probability of $1-\lambda$ that
$\theta$ does not equal $\theta_L$ is distributed uniformly over a given interval containing
$\theta_L$. Nevertheless, in the analysis of Bartlett's paradox that will now be presented, we will
retain all the assumptions that were made in the last section with regard to not only the sampling
density $g(x\,|\,\theta)$, but also with regard to the prior density of $\theta$ and the nature of
the interval $[\theta_L, \theta_U]$, since, by doing so, Bartlett's paradox becomes more relevant
to a real-world scenario without any of the essence of this paradox being~lost.

With the assumptions just clarified in place, Bartlett's paradox is the observation that the
posterior probability that the parameter $\theta$ lies in the interval $[-\varepsilon,
\varepsilon]$ tends to one as the parameter $\sigma_0$ of the prior density of $\theta$ given in
equation~(\ref{equ3}), which controls the variance of this density function, tends to infinity,
independent of what data set $x$ is observed.
Bartlett's paradox can only be really considered to be a paradox if, having established that there
was no or little pre-data information about the parameter $\theta$, one accepts the commonly-made
fallacy that placing a diffuse prior distribution over a parameter of interest is an adequate means
by which a lack of pre-data knowledge about the parameter may be represented.
A way in which we may try to avoid going down the road that leads to what Bartlett described in his
commentary as the `silly' limiting outcome that defines this paradox is, of course, to use any
pre-data knowledge that we may have about the parameter $\theta$ to elicit a value to $\sigma_0$
that is not too large.
However, in practice, we are still faced with the inconvenience that the posterior probability that
$\theta$ lies in the interval $[-\varepsilon, \varepsilon]$ will often be quite sensitive to the
choice made for $\sigma_0$.

To illustrate this point, rows~1 to~3 of Tables~\ref{tab1} to~\ref{tab3} show posterior
probabilities of $\theta$ lying in the interval $[-\varepsilon, \varepsilon]$ that result from
implementing the Bayesian method outlined in the previous section, which is labelled the `pure
Bayesian' method, in various different scenarios.
More precisely, these posterior probabilities have been derived under the assumption that the prior
probability that $\theta$ lies in the interval $[-\varepsilon, \varepsilon]$, i.e.\ the probability
$\lambda$, is equal to 0.4, the parameter $\theta_0$ of the prior density of $\theta$ given in
equation~(\ref{equ3}), which controls the location of this density function, is equal to zero, and
the standard error of the sample mean, i.e.\ $\sigma/\sqrt{n}$, is equal to one.
The assumptions made to derive the results presented in Tables~\ref{tab1} to~\ref{tab3} differ due
to the fact that these results correspond to the sample mean $\bar{x}$ being equal to 2.576, 0.8326
and zero, respectively.
To understand these choices for $\bar{x}$, we may take the value 2.576, which is the 0.995 quantile
of a standard normal density, as representing an extreme value for $\bar{x}$ under the null
hypothesis that $\theta=0$, while the value 0.8326 may be taken to represent a more `typical' value
for $\bar{x}$ under this hypothesis, however the particular reason for why this value was chosen
will become clear later.

\begin{table}[p!]
\caption{{\small Analysis of Bartlett's paradox part 1. Post-data probabilities of $\theta$ lying
in the interval $[-\varepsilon, \varepsilon]$ calculated assuming that $\lambda=0.4$, $\theta_0=0$,
$\sigma/\sqrt{n}=1$ and $\bar{x}=2.576$}}
\label{tab1}
\vspace{2ex}
{\renewcommand{\arraystretch}{1.18}
\hspace{-0.16in}
{\small
\begin{tabular}{|c|c|c|c|c|c|c|c|c|c|}
\hline
\multirow{2}{*}{Method} & \multirow{2}{*}{$\varepsilon$} &
\multicolumn{8}{c|}{Pre-data hyperparameter $\sigma_0$}\\
\cline{3-10}
& & 1 & 2 & 4 & 10 & 25 & 100 & 1000 & $\lim\hspace{-0.25em} \to\hspace{-0.25em} \infty$\\
\hline
\multirow{3}{*}{Pure Bayesian}
& 0 & 0.1522 & 0.0949 & 0.1080 & 0.2005 & 0.3779 & 0.7073 & 0.9602 & 1\\
& 0.1 & 0.1449 & 0.0923 & 0.1067 & 0.2003 & 0.3786 & 0.7084 & 0.9605 & 1\\
& 0.2 & 0.1387 & 0.0904 & 0.1061 & 0.2011 & 0.3808 & 0.7108 & 0.9609 & 1\\
\hline
\multirow{3}{*}{Fiducial-Bayes}
& 0 & 0.0489 & 0.0337 & 0.0327 & 0.0330 & 0.0330 & 0.0330 & 0.0330 & 0.0330\\
& 0.1 & 0.0489 & 0.0340 & 0.0330 & 0.0333 & 0.0333 & 0.0333 & 0.0333 & 0.0333\\
& 0.2 & 0.0520 & 0.0356 & 0.0344 & 0.0346 & 0.0346 & 0.0346 & 0.0346 & 0.0346\\
\hline
Mixture ($\kappa=0.2$)
& 0 & 0.0696 & 0.0459 & 0.0478 & 0.0665 & 0.1020 & 0.1679 & 0.2185 & 0.2264\\
\hline
\end{tabular}}}
\vspace{1ex}

\caption{{\small Analysis of Bartlett's paradox part 2. Post-data probabilities of $\theta$ lying
in the interval $[-\varepsilon, \varepsilon]$ calculated assuming that $\lambda=0.4$, $\theta_0=0$,
$\sigma/\sqrt{n}=1$ and $\bar{x}=0.8326$}}
\label{tab2}
\vspace{2ex}
{\renewcommand{\arraystretch}{1.18}
\hspace{-0.16in}
{\small
\begin{tabular}{|c|c|c|c|c|c|c|c|c|c|}
\hline
\multirow{2}{*}{Method} & \multirow{2}{*}{$\varepsilon$} &
\multicolumn{8}{c|}{Pre-data hyperparameter $\sigma_0$}\\
\cline{3-10}
& & 1 & 2 & 4 & 10 & 25 & 100 & 1000 & $\lim\hspace{-0.25em} \to\hspace{-0.25em} \infty$\\
\hline
\multirow{3}{*}{Pure Bayesian}
& 0 & 0.4422 & 0.5305 & 0.6648 & 0.8262 & 0.9219 & 0.9792 & 0.9979 & 1\\
& 0.1 & 0.4462 & 0.5377 & 0.6739 & 0.8333 & 0.9257 & 0.9804 & 0.9980 & 1\\
& 0.2 & 0.4508 & 0.5456 & 0.6834 & 0.8406 & 0.9296 & 0.9815 & 0.9981 & 1\\
\hline
\multirow{3}{*}{Fiducial-Bayes}
& 0 & 0.3795 & 0.3893 & 0.3966 & 0.3994 & 0.3999 & 0.4000 & 0.4000 & 0.4000\\
& 0.1 & 0.3777 & 0.3887 & 0.3966 & 0.3996 & 0.4001 & 0.4002 & 0.4002 & 0.4002\\
& 0.2 & 0.3764 & 0.3887 & 0.3971 & 0.4003 & 0.4009 & 0.4010 & 0.4010 & 0.4010\\
\hline
Mixture ($\kappa=0.2$)
& 0 & 0.3921 & 0.4175 & 0.4502 & 0.4848 & 0.5043 & 0.5158 & 0.5196 & 0.5200\\
\hline
\end{tabular}}}
\vspace{1ex}

\caption{{\small Analysis of Bartlett's paradox part 3. Post-data probabilities of $\theta$ lying
in the interval $[-\varepsilon, \varepsilon]$ calculated assuming that $\lambda=0.4$, $\theta_0=0$,
$\sigma/\sqrt{n}=1$ and $\bar{x}=0$}}
\label{tab3}
\vspace{2ex}
{\renewcommand{\arraystretch}{1.18}
\hspace{-0.16in}
{\small
\begin{tabular}{|c|c|c|c|c|c|c|c|c|c|}
\hline
\multirow{2}{*}{Method} & \multirow{2}{*}{$\varepsilon$} &
\multicolumn{8}{c|}{Pre-data hyperparameter $\sigma_0$}\\
\cline{3-10}
& & 1 & 2 & 4 & 10 & 25 & 100 & 1000 & $\lim\hspace{-0.25em} \to\hspace{-0.25em} \infty$\\
\hline
\multirow{3}{*}{Pure Bayesian}
& 0 & 0.4853 & 0.5985 & 0.7332 & 0.8701 & 0.9434 & 0.9852 & 0.9985 & 1\\
& 0.1 & 0.4942 & 0.6109 & 0.7457 & 0.8784 & 0.9475 & 0.9864 & 0.9986 & 1\\
& 0.2 & 0.5042 & 0.6244 & 0.7588 & 0.8867 & 0.9516 & 0.9875 & 0.9987 & 1\\
\hline
\multirow{3}{*}{Fiducial-Bayes}
& 0 & 0.4495 & 0.4721 & 0.4816 & 0.4847 & 0.4852 & 0.4853 & 0.4853 & 0.4853\\
& 0.1 & 0.4564 & 0.4804 & 0.4903 & 0.4935 & 0.4941 & 0.4942 & 0.4942 & 0.4942\\
& 0.2 & 0.4645 & 0.4899 & 0.5003 & 0.5037 & 0.5042 & 0.5043 & 0.5044 & 0.5044\\
\hline
Mixture ($\kappa=0.2$)
& 0 & 0.4566 & 0.4974 & 0.5319 & 0.5618 & 0.5768 & 0.5853 & 0.5879 & 0.5882\\
\hline
\end{tabular}}}
\end{table}

Under the assumptions that have just been made, it can be seen, from the rows in Tables~\ref{tab1}
to~\ref{tab3} that correspond to the use of the pure Bayesian method, how the posterior probability
of $\theta$ lying in the interval $[-\varepsilon, \varepsilon]$ that is defined by
equation~(\ref{equ21}) varies as the width $2\hspace{0.05em}\varepsilon$ of this interval goes from
zero to 0.4 and the hyperparameter $\sigma_0$ goes from one to infinity. In particular, the
information in these tables supports the conclusion that the posterior probability in question may
be generally regarded as being quite sensitive to changes in the value of $\sigma_0$.

\vspace{3ex}
\subsection{Lindley's paradox}
\label{sec16}

Lindley's paradox, as it has become known, is presented in Lindley~(1957). Variations on this
paradox are often referred to as examples of the Jeffreys-Lindley paradox due to Harold Jeffreys'
earlier work on related ideas.
In analysing Lindley's paradox, the same general assumptions will be made as were made when
analysing Bartlett's paradox in the previous section, i.e.\ the assumptions that originally were
made in Section~\ref{sec5} with regard to the sampling density $g(x\,|\,\theta)$, the prior density
of $\theta$ and the nature of the interval $[\theta_L, \theta_U]$. Nevertheless, the same
assumption will be made as in Lindley~(1957) with regard to how the observed sample mean $\bar{x}$
depends on the sample size $n$.
In particular, $\bar{x}$ will be assumed to be equal to $\Phi^{-1}(1-(\alpha/2))\sigma/\sqrt{n}$,
where $\Phi$ is the standard normal (cumulative) distribution function and $\alpha \in (0,1)$ is a
given constant, or in other words, $\bar{x}$ will be assumed to always fall exactly on the
$1-(\alpha/2)$ quantile of the sampling density of $\bar{x}$ under the null hypothesis that
$\theta$ is equal to zero.

In the context of the assumptions that have just been made, Lindley's paradox is, in essence, the
observation that, in the case where $\varepsilon=0$, the posterior probability that the parameter
$\theta$ lies in the interval $[-\varepsilon, \varepsilon]$, or in other words equals 0, tends to
one as the sample size $n$ tends to infinity, independent of what value is assigned to $\alpha$ in
the definition of the observed mean $\bar{x}$.
Under the usual interpretation of Lindley's paradox, the fact that $\alpha$ is the two-sided P
value that results from testing the null hypothesis that $\theta$ equals zero is used as a
justification for treating $\alpha$ as being a measure of the compatibility of this null hypothesis
with the observed mean $\bar{x}$. However, as explained in Section~7 of Bowater~(2024),
interpreting a P value in this way is a fallacy. For this reason, let us use a more sensible
interpretation of the observation that defines Lindley's paradox that still justifies why it may be
regarded as being a paradox.

In particular, let us begin by noting that the expected height of the standard normal density at a
value that is randomly drawn from this density function is $1/(2\sqrt{\pi}\hspace{0.07em})=0.2821$.
By contrast, the height of the standard normal density at its 0.9995 quantile is only 0.0018, i.e.\
only 0.63\% of its expected height as just defined. For this reason, if the sample mean $\bar{x}$
falls on the 0.9995 quantile of the sampling density for $\bar{x}$ under the null hypothesis that
$\theta=0$, then we can justify describing this hypothesis as being not well supported by the data.

However, even though the null hypothesis that $\theta=0$ is not well supported by the data, perhaps
our pre-data knowledge about the parameter $\theta$ may justify placing a large post-data
probability on this hypothesis being true? In responding to this question, we may first note that
the observation that defines Lindley's paradox holds no matter what value $\lambda$ is assigned to
the prior probability that $\theta$ is equal to zero as long as this value is positive. Second, for
very large $n$, it could be reasonably expected that in almost all practical scenarios in which we
may find ourselves, our pre-data knowledge about $\theta$ when $\theta$ is conditioned not to equal
zero would be heavily or completely swamped by the large amount of data that has been observed.
However, according to the limiting outcome that defines Lindley's paradox, this clearly does not
happen when the Bayesian method outlined in Section~\ref{sec5} is applied to the problem of
inference under discussion. Therefore, this can be identified as being a paradox.

In summary, Lindley's paradox helps us to appreciate that although the Bayesian analogy may be
useful in a given scenario under certain conditions, if these conditions change, it may be regarded
as having been stretched beyond the point of being useful. To be more precise, while the Bayesian
analogy that underlies the method of inference being discussed may be considered to be useful when
the sample size $n$ is small, increasing $n$ will arguably, in most practical scenarios, push the
Bayesian analogy towards a breaking point, beyond which it is no longer considered to be useful.

We should point out that, although Lindley's paradox only directly applies to the case where
$\varepsilon$ is equal to zero, its distorting effect on the posterior probability that the
parameter $\theta$ lies in the interval $[-\varepsilon, \varepsilon]$ as the sample size $n$
becomes larger may be appreciated for values of $\varepsilon$ that are greater than zero.
Indeed, for small to moderate values of $n$ and for small positive values of $\varepsilon$, the
distance between the observed sample mean $\bar{x}$ and the limits of the interval $[-\varepsilon,
\varepsilon]$ will be almost equal to $\bar{x}$, and so the effect of increasing $n$ on the
posterior probability of the parameter $\theta$ lying in the interval $[-\varepsilon, \varepsilon]$
will be similar to the case in which $\varepsilon=0$.
However, if the sample size $n$ continues to be increased after a certain level has been passed,
then it will become unclear whether the tendency for the posterior probability in question to
increase towards one is overwhelming due to the effect of Lindley's paradox, or whether this
tendency is also substantially due to the fact that the sample mean $\bar{x}$ is getting relatively
much closer to the values in the upper half of the interval $[-\varepsilon, \varepsilon]$ than to
the value at the centre of this interval, i.e.\ the value zero.

To illustrate the points that have just been made, rows~1 to~3 of Tables~\ref{tab4} and~\ref{tab5}
show posterior probabilities of $\theta$ lying in the interval $[-\varepsilon, \varepsilon]$ that
result from implementing the Bayesian method under discussion, i.e.\ the `pure Bayesian' method, in
various different scenarios.
In common with the results given in Tables~\ref{tab1} to~\ref{tab3}, the results presented in these
two tables have been derived under the assumption that the prior probability that $\theta$ lies in
the interval $[-\varepsilon,\varepsilon]$, i.e.\ the probability $\lambda$, is equal to 0.4.
The results given in both Tables~\ref{tab4} and~\ref{tab5} correspond, furthermore, to the
population standard deviation $\sigma$ being equal to 4 and to the observed sample mean $\bar{x}$
always falling on the 0.995 quantile of the sampling density of $\bar{x}$ under the null hypothesis
that $\theta$ equals zero, i.e.\ to $\alpha$ being equal to 0.01, which is a property on which the
results presented in Table~\ref{tab1} were also derived.
The assumptions made to derive the results presented in the rows in Tables~\ref{tab4}
and~\ref{tab5} that correspond to the use of the pure Bayesian method differ between these two
tables due to the fact that the results of interest in Table~\ref{tab4} correspond to the
parameters $\theta_0$ and $\sigma_0$ of the prior density of $\theta$ given in
equation~(\ref{equ3}) being equal to 0 and 4, respectively, which is consistent with this prior
density being `mildly informative' outside of the interval $[-\varepsilon, \varepsilon]$, while the
results of interest in Table~\ref{tab5} correspond to these parameters being equal to 1.5 and one,
respectively, which is consistent with the prior density of $\theta$ being `moderately informative'
outside of the interval $[-\varepsilon, \varepsilon]$.

\begin{table}[t!]
\caption{{\small Analysis of Lindley's paradox part 1. Post-data probabilities of $\theta$ lying in
the interval $[-\varepsilon, \varepsilon]$ calculated assuming that $\lambda=0.4$, $\theta_0=0$,
$\sigma_0=4$, $\sigma=4$ and $\bar{x}=\Phi^{-1}(0.995)\sigma/\sqrt{n}$}}
\label{tab4}
\vspace{2ex}
{\renewcommand{\arraystretch}{1.18}
\hspace{-0.49in}
{\small
\begin{tabular}{|c|c|c|c|c|c|c|c|c|c|c|}
\hline
\multirow{2}{*}{Method} & \multirow{2}{*}{$\varepsilon$} &
\multicolumn{9}{c|}{Sample size $n$}\\
\cline{3-11}
& & 1 & 4 & 10 & 20 & 50 & 200 & 1000 & 5000 & $\lim\hspace{-0.25em} \to\hspace{-0.25em} \infty$\\
\hline
\multirow{3}{*}{Pure Bayesian}
& 0 & 0.1522 & 0.0949 & 0.0977 & 0.1148 & 0.1556 & 0.2582 & 0.4340 & 0.6309 & 1\\
& 0.05 & 0.1512 & 0.0942 & 0.0970 & 0.1140 & 0.1548 & 0.2593 & 0.4469 & 0.6902 & 1\\
& 0.1 & 0.1503 & 0.0935 & 0.0964 & 0.1135 & 0.1549 & 0.2645 & 0.4859 & 0.8160 & 1\\
\hline
\multirow{3}{*}{Fiducial-Bayes}
& 0 & 0.0489 & 0.0337 & 0.0327 & 0.0327 & 0.0329 & 0.0330 & 0.0330 & 0.0330 & 0.0330\\
& 0.05 & 0.0488 & 0.0337 & 0.0328 & 0.0328 & 0.0331 & 0.0342 & 0.0411 & 0.0840 & 1\\
& 0.1 & 0.0487 & 0.0337 & 0.0329 & 0.0332 & 0.0341 & 0.0392 & 0.0729 & 0.2461 & 1\\
\hline
Mixture ($\kappa=0.2$)
& 0 & 0.0696 & 0.0459 & 0.0457 & 0.0492 & 0.0574 & 0.0780 & 0.1132 & 0.1526 & 0.2264\\
\hline
\end{tabular}}}
\vspace{1ex}

\caption{{\small Analysis of Lindley's paradox part 2. Post-data probabilities of $\theta$ lying in
the interval $[-\varepsilon, \varepsilon]$ calculated assuming that $\lambda=0.4$, $\theta_0=1.5$,
$\sigma_0=1$, $\sigma=4$ and $\bar{x}=\Phi^{-1}(0.995)\sigma/\sqrt{n}$}}
\label{tab5}
\vspace{2ex}
{\renewcommand{\arraystretch}{1.18}
\hspace{-0.49in}
{\small
\begin{tabular}{|c|c|c|c|c|c|c|c|c|c|c|}
\hline
\multirow{2}{*}{Method} & \multirow{2}{*}{$\varepsilon$} &
\multicolumn{9}{c|}{Sample size $n$}\\
\cline{3-11}
& & 1 & 4 & 10 & 20 & 50 & 200 & 1000 & 5000 & $\lim\hspace{-0.25em} \to\hspace{-0.25em} \infty$\\
\hline
\multirow{3}{*}{Pure Bayesian}
& 0 & 0.1957 & 0.0930 & 0.0529 & 0.0416 & 0.0468 & 0.1047 & 0.2750 & 0.5162 & 1\\
& 0.05 & 0.1945 & 0.0921 & 0.0523 & 0.0412 & 0.0465 & 0.1050 & 0.2859 & 0.5839 & 1\\
& 0.1 & 0.1933 & 0.0914 & 0.0519 & 0.0410 & 0.0465 & 0.1081 & 0.3226 & 0.7410 & 1\\
\hline
\multirow{3}{*}{Fiducial-Bayes}
& 0 & 0.1586 & 0.0605 & 0.0357 & 0.0307 & 0.0310 & 0.0328 & 0.0331 & 0.0331 & 0.0330\\
& 0.05 & 0.1582 & 0.0604 & 0.0358 & 0.0309 & 0.0313 & 0.0340 & 0.0406 & 0.0821 & 1\\
& 0.1 & 0.1580 & 0.0606 & 0.0360 & 0.0313 & 0.0323 & 0.0385 & 0.0699 & 0.2420 & 1\\
\hline
Mixture ($\kappa=0.2$)
& 0 & 0.1661 & 0.0670 & 0.0391 & 0.0329 & 0.0342 & 0.0471 & 0.0815 & 0.1297 & 0.2264\\
\hline
\end{tabular}}}
\vspace{3.5ex}
\end{table}

Under the assumptions that have just been made, it can be seen, from the rows in Tables~\ref{tab4}
and~\ref{tab5} that correspond to the use of the pure Bayesian method, how the posterior
probability of $\theta$ lying in the interval $[-\varepsilon, \varepsilon]$ that is defined by
equation~(\ref{equ21}) varies as the width $2\hspace{0.05em}\varepsilon$ of this interval goes from
zero to 0.2 and the sample size $n$ goes from one to infinity.
More specifically, in all the scenarios analysed in the rows of Tables~\ref{tab4} and~\ref{tab5}
being referred to, it may be appreciated that, as the sample size $n$ increases, the posterior
probability of $\theta$ lying in the interval $[-\varepsilon, \varepsilon]$ initially decreases and
then having reached a minimum, increases towards one.
To give some more detail, it may be pointed out that when $\varepsilon$ is equal to zero, the
minimum value for the posterior probability under discussion is equal to 0.0931 and occurs at $n=6$
for the case analysed in Table~\ref{tab4}, and is equal to 0.0408 and occurs at $n=25$ for the case
analysed in Table~\ref{tab5}.

Also, from looking at rows~1 to~3 of Tables~\ref{tab4} and~\ref{tab5}, it may be noticed that, in
the case where $n=5000$, the posterior probability of $\theta$ lying in the interval
$[-\varepsilon, \varepsilon]$ is much larger when $\varepsilon$ is equal to 0.05 and 0.1 than when
it is equal to zero.
However, the reason for this should not be attributed to the effect of Lindley's paradox, but
rather to the fact that the sample mean $\bar{x}$ is equal to 0.1457 in the case where $n=5000$,
and is therefore, relatively much closer to the two positive values of $\varepsilon$ in question
than to zero, i.e.\ the reason should be attributed to the effect of a phenomenon that was
discussed earlier.

\vspace{3ex}
\section{Example of the combined use of three different types of analogy}
\label{sec22}

\vspace{1.5ex}
\subsection{The analogies that will be used}
\label{sec14}

We will now address the reference problem of inference that was outlined in Section~\ref{sec4} by
combining three different types of analogy, namely the fiducial analogy, the Bayesian analogy and
an artificial data analogy. In addressing this problem, it will again be assumed that the interval
$[\theta_L, \theta_U]$ is the interval $[-\varepsilon, \varepsilon]$, where $\varepsilon \geq 0$.

To begin with, let us clarify that, as was the case for the Bayesian method of inference presented
in Section~\ref{sec5}, the method of inference that will now be developed is based on the global
analogy that was introduced by Assumption~\ref{asm2} in Section~\ref{sec4}. However, in contrast to
the analogy that underlies the Bayesian method in question, it is based on the special case of this
global analogy that corresponds to neither the composition of the subset of the population to be
randomly sampled that only contains values of the parameter $\theta$ lying outside of the interval
$[-\varepsilon, \varepsilon]$, nor the composition of the subset of the population to be sampled
that only contains values of $\theta$ lying inside this interval being completely known.
This special case of the global analogy being referred to is also a special case of the pre-data
parameter sampling analogy that was labelled as Example~\ref{exa3} in Section~\ref{sec6}.
In particular, it is the special case of this latter analogy in which there are only two possible
populations of values for the parameter $\theta$, i.e.\ the populations $P_1$ and $P_2$, where the
population $P_1$ only contains values of $\theta$ that lie outside of the interval $[-\varepsilon,
\varepsilon]$ and the population $P_2$ only contains values of $\theta$ that lie inside this
interval, and where there is a known (physical) probability of $\lambda$ that population $P_2$ will
be selected as the population to be sampled.

\vspace{3ex}
\subsection{Conditional use of the fiducial analogy}
\label{sec9}

To develop the method of inference under current consideration, we will begin by assuming that,
apart from the aforementioned restrictions on the values of the parameter $\theta$ that may lie in
what were just defined as the populations $P_1$ and $P_2$, the composition of each of these two
populations is completely unknown.
As was discussed in Section~\ref{sec6}, if an analogy is to be used for expressing pre-data
knowledge about a parameter of interest in which the true value of this parameter is about to be
randomly drawn from a population of possible values for the parameter, then in order for the
analogy to be consistent with the scenario of interest being ideal for the application of the
fiducial analogy, the composition of this population would need to be completely unknown.
Nevertheless, in the case that is presently of concern, the use of the fiducial analogy for making
post-data inferences about the parameter $\theta$ conditional on $\theta$ lying inside the interval
$[-\varepsilon, \varepsilon]$ or conditional on $\theta$ lying outside of this interval is not
appropriate due to the limits placed on the natural range of $\theta$ by both of these conditions.

It is possible though, as an alternative, to make post-data inferences about the parameter $\theta$
under either of these two conditions by using a slightly modified version of the fiducial analogy.
More specifically, given what has already been assumed and using terminology introduced in
Section~\ref{sec2}, it would seem quite acceptable that, instead of assuming that the distribution
of an appropriately chosen primary r.v.\ $\Gamma$ does not change on observing the data $x$, it is
only assumed that, over values of the variable $\Gamma$ that are possible after the data $x$ have
been observed, the relative height of the density function of $\Gamma$ does not change on observing
the data.
This modified argument concerning what should be the post-data distribution of the primary r.v.\ is
referred to as the moderate fiducial argument in Bowater~(2019, 2022b) where further discussion of
this argument may be found.

Under the assumption that the sampling density $g(x_i\,|\,\theta)$ of each independent data value
$x_i$ is a normal density with mean $\theta$ and known variance $\sigma^2$, and the primary r.v.\
$\Gamma$ is as defined in Section~\ref{sec2}, it is trivial to show that if the moderate fiducial
argument was applied to the problem of inference being considered under the condition that $\theta$
lies in the interval $[-\varepsilon, \varepsilon]$, then the post-data or fiducial density of
$\theta$ would be given by the expression:
\vspace{1ex}
\begin{equation}
\label{equ8}
f(\theta\,|\,\theta \in [-\varepsilon, \varepsilon],x) = \left\{
\begin{array}{ll}
\mathtt{C}_2\hspace{0.05em} \phi((\theta-\bar{x})\sqrt{n}/ \sigma)\, & \mbox{if $\theta \in
[-\varepsilon, \varepsilon]$}\\[1ex]
0 & \mbox{otherwise}
\end{array}
\right.
\vspace{1ex}
\end{equation}
where $\mathtt{C}_2$ is a normalising constant, while if this argument was applied under the
con\-di\-tion that $\theta$ does not lie in the interval $[-\varepsilon, \varepsilon]$, then the
fiducial density of $\theta$ would be given by the expression:
\vspace{0.5ex}
\begin{equation}
\label{equ9}
f(\theta\,|\,\theta \notin [-\varepsilon, \varepsilon],x) = \left\{
\begin{array}{ll}
\mathtt{C}_3\hspace{0.05em} \phi((\theta-\bar{x})\sqrt{n}/ \sigma)\, & \mbox{if $\theta \notin
[-\varepsilon, \varepsilon]$}\\[1ex]
0 & \mbox{otherwise}
\end{array}
\right.
\vspace{1ex}
\end{equation}
where $\mathtt{C}_3$ is a normalising constant.

\vspace{3ex}
\subsection{Conditional use of an artificial data analogy}
\label{sec10}

Let us now consider combining the fiducial analogy with an analogy of the type referred to in
Section~\ref{sec7}, namely an artificial data analogy, and a concept related to this latter type of
analogy that was discussed in this earlier section, namely a GPD function of the parameter of
interest.
In particular, let us apply the fiducial analogy to the reference problem of inference described in
Section~\ref{sec4} under the assumption that the GPD function of the parameter $\theta$ is either
defined by the expression:
\vspace{1ex}
\begin{equation}
\label{equ4}
\omega_G(\theta) =
\left\{
\begin{array}{ll}
\phi((\theta-\theta_0)/ \sigma_0)(1+\tau h(\theta))\, & \mbox{if $\theta \in [-\varepsilon,
\varepsilon]$}\\[1ex]
0 & \mbox{otherwise}
\end{array}
\right.
\vspace{1ex}
\end{equation}
or by the expression:
\begin{equation}
\label{equ5}
\omega_G(\theta) =
\left\{
\begin{array}{ll}
0 & \mbox{if $\theta \in [-\varepsilon, \varepsilon]$}\\[1ex]
\phi((\theta-\theta_0)/ \sigma_0)\, & \mbox{otherwise}
\end{array}
\right.
\vspace{1.5ex}
\end{equation}
where $\theta_0\hspace{-0.05em} \in\hspace{-0.05em} (-\infty, \infty)$, $\sigma_0>0$ and
$\tau \geq 0$ are given constants and the density function\linebreak $h(\theta)$ is as defined in
Section~\ref{sec5}.
In doing this, the fiducial density of $\theta$ that is derived using the GPD function defined by
equation~(\ref{equ4}) will be assumed to be the fiducial density of $\theta$ conditional on
$\theta$ lying inside the interval $[-\varepsilon, \varepsilon]$, which we will denote as the
fiducial density $f(\theta\,|\,\theta \in [-\varepsilon, \varepsilon],x)$, while the fiducial
density of $\theta$ that is derived using the GPD\linebreak function defined by
equation~(\ref{equ5}) will be assumed to be the fiducial density of $\theta$ conditional on
$\theta$ lying outside of the interval $[-\varepsilon, \varepsilon]$, which we will denote as the
fiducial density $f(\theta\,\hspace{0.05em}|\,\hspace{0.05em}\theta\hspace{-0.05em}
\notin\hspace{-0.05em} [-\varepsilon, \varepsilon],x)$.
Under these assumptions and those made in the previous section with regard to the sampling density
$g(x\,|\,\theta)$ and the primary r.v.\ $\Gamma$, the conditional fiducial densities in question
are given by the following expressions:
\vspace{1.25ex}
\begin{equation}
\label{equ10}
f(\theta\,|\,\theta \in [-\varepsilon, \varepsilon],x) =
\left\{
\begin{array}{ll}
\mathtt{C}_4\hspace{0.05em} \phi((\theta-\theta_1)/ \sigma_1)(1+\tau h(\theta))\, & \mbox{if
$\theta \in [-\varepsilon, \varepsilon]$}\\[1ex]
0 & \mbox{otherwise}
\end{array}
\right.
\vspace{1ex}
\end{equation}
\begin{equation}
\label{equ11}
f(\theta\,|\,\theta \notin [-\varepsilon, \varepsilon],x) =
\left\{
\begin{array}{ll}
\mathtt{C}_5\hspace{0.05em} \phi((\theta-\theta_1)/ \sigma_1)\, & \mbox{if $\theta \notin
[-\varepsilon, \varepsilon]$}\\[1ex]
0 & \mbox{otherwise}
\end{array}
\right.
\vspace{1.25ex}
\end{equation}
where $\mathtt{C}_4$ and $\mathtt{C}_5$ are normalising constants and where $\theta_1$ and
$\sigma_1$ are defined by the following expressions:
\vspace{1ex}
\begin{equation}
\label{equ18}
\theta_1 = \frac{\sigma_0^2\bar{x} + (\sigma^2\hspace{-0.1em}/n)\theta_0}{\sigma_0^2 +
(\sigma^2\hspace{-0.1em}/n)}
\vspace{0.5ex}
\end{equation}
\begin{equation}
\label{equ19}
\sigma_1^2 = \frac{(\sigma^2\hspace{-0.1em}/n)\sigma_0^2}{(\sigma^2\hspace{-0.1em}/n)+\sigma_0^2}
\vspace{1ex}
\end{equation}

Observe that, if $(\sigma/ \sigma_0)^2$ is a whole number, then the way that the fiducial density
$f(\hspace{0.05em}\theta\,\hspace{0.1em}|\,\hspace{0.1em}\theta\hspace{-0.15em}
\notin\hspace{-0.15em} [-\varepsilon, \varepsilon],x\hspace{0.05em})$ has been derived is
equivalent to applying the moderate fiducial argument to the problem of inference being discussed
under the condition that $\theta$ lies outside of the interval $[-\varepsilon, \varepsilon]$ after
having observed not only the data set $x$, but also a preliminary data set $y$ that has a mean of
$\theta_0$ and a sample size equal to $(\sigma/ \sigma_0)^2$.
In other words, the way that the fiducial density $f(\theta\,|\,\theta \notin [-\varepsilon,
\varepsilon],x)$ has been derived by using the GPD function of $\theta$ defined by
equation~(\ref{equ5}) is equivalent to using an artificial data analogy and the slightly modified
form of the fiducial analogy that was described in the preceding section.

\vspace{3ex}
\subsection{A limited type of Bayesian analogy}
\label{sec12}

We will now turn our attention to addressing the reference problem of inference described in
Section~\ref{sec4} in the case that is of principal interest, i.e.\ the case where the parameter
$\theta$ is not conditioned to lie in any particular region of the real line.
In pursuing this goal, we will use a Bayesian analogy, but one that is different from the type of
Bayesian analogy that was used in Section~\ref{sec5}.

In particular, we will make an analogy between our pre-data knowledge about the parameter $\theta$
and what would be our uncertainty about the true value of this parameter if it was about to be
randomly drawn from a population of values for $\theta$, each member of which can only be equal to
one of two specific values for $\theta$, namely $\theta_A$ and $\theta_B$, where $\theta_A$ lies in
the interval $[-\varepsilon, \varepsilon]$, while $\theta_B$ lies outside of this interval.
The proportion of values in the population in question that are equal to $\theta_A$ will be assumed
to be equal to the value $\lambda$ as this value was defined in Section~\ref{sec4}.
Also, it will be assumed that, as an expression of our pre-data knowledge about the parameter
$\theta$, this analogy will be maintained active after the data $x$ have been observed.
Moreover, the key detail that establishes this analogy as an analogy that possesses a substantial
amount of practical relevance is that, apart from it being known that $\theta_A$ and $\theta_B$ lie
inside and outside of the interval $[-\varepsilon, \varepsilon]$ respectively, these values will be
assumed to be unknown, i.e.\ completely or partially unknown.
Observe that this analogy is, in effect, a special case of the global analogy introduced by
Assumption~\ref{asm2} in Section~\ref{sec4}.

Let us, though, for a brief moment, imagine that the final assumption that was just made is
withdrawn and the values of $\theta_A$ and $\theta_B$ are in fact known.
In this case, by means of a straightforward application of Bayes' theorem, the posterior
probabilities of the parameter $\theta$ being equal to $\theta_A$ and of it being equal to
$\theta_B$ would be given, respectively, by the following two expressions:
\begin{equation}
\label{equ6}
P_0(\theta = \theta_{A}\,|\,x) =
\frac{\lambda g(x\,|\,\theta_{A})}{\lambda g(x\,|\,\theta_{A}) + (1-\lambda)g(x\,|\,\theta_{B})}
\vspace{0.5ex}
\end{equation}
\begin{equation}
\label{equ7}
P_0(\theta = \theta_{B}\,|\,x) = 1 - P_0(\theta = \theta_{A}\,|\,x)
\end{equation}
Note that these expressions are valid for any sampling density $g(x\,|\,\theta)$ that depends on
the true value of a given unknown parameter $\theta$ and not just a normal sampling density that
depends on the true value of the unknown population mean $\theta$.
We will maintain this level of generality concerning the definition of the sampling density
$g(x\,|\,\theta)$ in the conceptual development that immediately follows.

Of course, if the values of $\theta_A$ and $\theta_B$ are unknown, then the height of the sampling
density $g(x\,|\,\theta)$ at the point where the observed data $x$ lies assuming that $\theta$ is
equal to $\theta_A$ or assuming that $\theta$ is equal to $\theta_B$ will be unknown, i.e.\ the
values $g(x\,|\,\theta_{A})$ and $g(x\,|\,\theta_{B})$ will be unknown.
Therefore, in this case, we may not directly use the expressions in equation~(\ref{equ6})
and~(\ref{equ7}) to determine the posterior probabilities of the parameter $\theta$ being equal to
$\theta_{A}$ and of it being equal to $\theta_{B}$, i.e.\ the posterior probabilities of $\theta$
lying inside the interval $[-\varepsilon, \varepsilon]$ and of it lying outside of this interval.

Nevertheless, it would seem reasonable, at least at first sight, to consider substituting the
quantities $g(x\,|\,\theta_{A})$ and $g(x\,|\,\theta_{B})$ in equation~(\ref{equ6}) by expected
values of these two quantities that are determined on the basis of all the information that is
available to us after the data $x$ have been observed.
To be precise, these expected values of $g(x\,|\,\theta_{A})$ and $g(x\,|\,\theta_{B})$ would be
determined with respect to post-data distributions of the values $\theta_{A}$ and $\theta_{B}$,
respectively, and to clarify even further, these expected values would be determined using the
following expressions:
\begin{equation}
\label{equ16}
\mathbb{E}[g(x\,|\,\theta_{A})] = \int_{-\varepsilon}^{\varepsilon} g(x\,|\,\theta_{A})
p(\theta_{A}\,|\,x) d \theta_{A}
\end{equation}
\begin{equation}
\label{equ17}
\mathbb{E}[g(x\,|\,\theta_{B})] = \int_{\varepsilon}^{\infty}\hspace{-0.1em} g(x\,|\,\theta_{B})
p(\theta_{B}\,|\,x) d \theta_{B} + \int_{-\infty}^{-\varepsilon} g(x\,|\,\theta_{B})
p(\theta_{B}\,|\,x) d \theta_{B}
\vspace{1ex}
\end{equation}
where $p(\theta_{A}\,|\,x)$ and $p(\theta_{B}\,|\,x)$ are post-data density functions of
$\theta_{A}$ and $\theta_{B}$, respectively.

An immediate issue that we are confronted with, though, in implementing this kind of strategy is
how do we determine the post-data densities $p(\theta_{A}\,|\,x)$ and $p(\theta_{B}\,|\,x)$ in a
way that is compatible with the limited type of Bayesian analogy that has already been made?
In effect, this means how do we determine these post-data densities in a way that takes into
account the discussion that was presented in Section~\ref{sec8} concerning the general need for
analogies to be compatible when multiple analogies are being applied to the same problem of
inference?

\vspace{3ex}
\subsection{Compatibility between the analogies being used}

To address the question that has just been raised, let us begin by assuming that with regard to the
case where the sampling density $g(x_i\,|\,\theta)$ of each independent data value $x_i$ is a
normal density with mean $\theta$ and known variance $\sigma^2$, the post-data densities
$p(\theta_{A}\,|\,x)$ and $p(\theta_{B}\,|\,x)$ are equivalent to what in equations~(\ref{equ8})
and~(\ref{equ9}) were defined as being the fiducial densities $f(\theta\,|\,\theta \in
[-\varepsilon, \varepsilon],x)$ and $f(\theta\,|\,\theta \notin [-\varepsilon, \varepsilon],x)$,
respectively.

As alluded to in Section~\ref{sec9}, if an analogy is to be used for expressing pre-data knowledge
about the parameter $\theta$ in which the true value of $\theta$ is about to be randomly drawn from
a population of possible values for $\theta$, then in order for the analogy to be consistent with
the scenario of interest being ideal for using the moderate fiducial argument to derive either of
the fiducial densities
$f(\hspace{0.05em}\theta\,\hspace{0.1em}|\,\hspace{0.1em}\theta\hspace{-0.15em} \in\hspace{-0.15em}
[-\varepsilon, \varepsilon],x\hspace{0.05em})$ or
$f(\hspace{0.05em}\theta\,\hspace{0.1em}|\,\hspace{0.1em}\theta\hspace{-0.15em}
\notin\hspace{-0.15em} [-\varepsilon, \varepsilon],x\hspace{0.05em})$, it would need to be the case
that this population, apart from containing values of $\theta$ that are either restricted to lie
inside the interval $[-\varepsilon, \varepsilon]$ or restricted to lie outside of this interval,
has a composition that is completely unknown.
Also, we may observe that this latter requirement is consistent with the value $\theta_A$, i.e.\
the value of $\theta$ under the condition that $\theta\hspace{-0.05em} \in\hspace{-0.05em}
[-\varepsilon, \varepsilon]$, and the value $\theta_B$, i.e.\ the value of $\theta$ under the
condition that $\theta \notin [-\varepsilon, \varepsilon]$, being simply fixed and completely
unknown values at any time point before the data set $x$ is observed, except for the aforementioned
restrictions on where on the real line these values may lie.
Therefore, in this regard, the slightly modified version of the fiducial analogy that was used to
derive the fiducial densities $f(\theta\,|\,\theta \in [-\varepsilon, \varepsilon],x)$ and
$f(\theta\,\hspace{0.05em}|\,\hspace{0.05em}\theta\hspace{-0.05em} \notin\hspace{-0.05em}
[-\varepsilon, \varepsilon],x)$ defined by equations~(\ref{equ8}) and~(\ref{equ9}) and the limited
type of Bayesian analogy that was introduced in the preceding section are compatible.
Furthermore, this type of compatibility is unaffected if instead of the fiducial densities
$f(\theta\,|\,\theta \in [-\varepsilon, \varepsilon],x)$ and $f(\theta\,|\,\theta \notin
[-\varepsilon, \varepsilon], x)$ being derived by only using the moderate fiducial argument, they
are derived by combining this argument with an artificial data analogy or with a GDP function of
the parameter $\theta$ in the way that was described in Section~\ref{sec10}, i.e.\ they are defined
according to the expressions in equations~(\ref{equ10}) and~(\ref{equ11}).

To introduce another area in which we may question the compatibility of the slightly modified
version of the fiducial analogy that is under discussion and the limited type of Bayesian analogy
that was put forward in the previous section, let us begin by pointing out that, in using Bayes'
theorem to make post-data inferences about a parameter of interest in any given situation, the
observed data set is only called upon once and that is when the joint distribution of the data and
the parameter of interest is conditioned on the observed data to obtain the posterior distribution
of the parameter.
By contrast, using the fiducial densities
$f(\hspace{0.05em}\theta\,\hspace{0.1em}|\,\hspace{0.1em}\theta\hspace{-0.15em} \in\hspace{-0.15em}
[-\varepsilon, \varepsilon],x\hspace{0.05em})$ and
$f(\hspace{0.05em}\theta\,\hspace{0.1em}|\,\hspace{0.1em}\theta\hspace{-0.15em}
\notin\hspace{-0.15em} [-\varepsilon, \varepsilon],x\hspace{0.05em})$ defined either by
equations~(\ref{equ8}) and~(\ref{equ9}) or by equations~(\ref{equ10}) and~(\ref{equ11}) to
determine expected values for the quantities $g(x\,|\,\theta_{A})$ and $g(x\,|\,\theta_{B})$ in
accordance with the strategy outlined in the last section implies that, in effect, the data are
employed for a second time after Bayes' theorem has already been applied.
Nevertheless, when we take into account the conditions under which this additional utilisation of
the observed data takes place, it becomes clear that this is not a source of incompatibility
between the versions of the fiducial and Bayesian analogies that are under consideration.

In particular, if the condition that the parameter $\theta$ must lie inside the interval
$[-\varepsilon, \varepsilon]$ or that it must lie outside of this interval is applied to the
limited type of Bayesian analogy introduced in the previous section, then this parameter must be
equal to $\theta_A$ or $\theta_B$, respectively, which must be known values in order to be able to
use the standard definitions of the posterior probabilities of $\theta$ being equal to $\theta_A$
and of it being equal to $\theta_B$ given in equations~(\ref{equ6}) and~(\ref{equ7}).
Under either of the conditions in question, nothing therefore can be learned from the data $x$
about the parameter $\theta$ in the scenario to which these standard definitions correspond because
the value of this parameter will be known, i.e.\ it will be known to be either equal to $\theta_A$
or $\theta_B$.
In other words, the observed data is irrelevant in the scenario being referred to if either the
condition that the parameter $\theta$ lies inside the interval $[-\varepsilon, \varepsilon]$ or
that it lies outside of this interval is enforced, which justifies why we may legitimately apply
the slightly modified version of the fiducial analogy being discussed to the observed data $x$
under either of these conditions in order to obtain the fiducial densities
$f(\hspace{0.05em}\theta\,\hspace{0.1em}|\,\hspace{0.1em}\theta\hspace{-0.15em} \in\hspace{-0.15em}
[-\varepsilon, \varepsilon],x\hspace{0.05em})$ and
$f(\hspace{0.05em}\theta\,\hspace{0.1em}|\,\hspace{0.1em}\theta\hspace{-0.15em}
\notin\hspace{-0.15em} [-\varepsilon, \varepsilon],x\hspace{0.05em})$ even when the Bayesian
analogy under consideration is also being applied to the problem of inference in question.

\vspace{3ex}
\subsection{Clarifying and illustrating the solution advocated}
\label{sec15}

Having hopefully removed some natural doubts about the legitimacy of addressing the reference
problem of inference outlined in Section~\ref{sec4} by using a limited type of Bayesian analogy, a
slightly modified version of the fiducial analogy, an artificial data analogy and a GPD function of
the parameter $\theta$ alongside each other in the way that was described in the previous two
sections, let us now clarify the solution to this problem of inference that results from
implementing the method in question.

In particular, we may obtain definitions of the post-data probabilities of the parameter $\theta$
lying in the interval $[-\varepsilon, \varepsilon]$ and of it not lying in this interval that are
in accordance with what was assumed in Sections~\ref{sec14} to~\ref{sec12} by re-expressing
equations~(\ref{equ6}) and~(\ref{equ7}) in the following way:
\begin{equation}
\label{equ12}
P_0(\theta \in [-\varepsilon, \varepsilon]\,|\,x) =
\frac{\lambda \mathbb{E}[g(x\,|\,\theta_{A})]}{\lambda \mathbb{E}[g(x\,|\,\theta_{A})] +
(1-\lambda)\mathbb{E}[g(x\,|\,\theta_{B})]}
\vspace{0.5ex}
\end{equation}
\begin{equation}
\label{equ13}
P_0(\theta \notin [-\varepsilon, \varepsilon]\,|\,x) =
1 - P_0(\theta \in [-\varepsilon, \varepsilon]\,|\,x)
\vspace{0.5ex}
\end{equation}
where $\mathbb{E}[g(x\,|\,\theta_{A})]$ and $\mathbb{E}[g(x\,|\,\theta_{B})]$ are determined
according to the expressions in equations~(\ref{equ16}) and~(\ref{equ17}) with the post-data
densities $p(\theta_{A}\,|\,x)$ and $p(\theta_{B}\,|\,x)$ in these expressions assumed to be,
respectively, the fiducial densities
$f(\hspace{0.07em}\theta\,\hspace{0.15em}|\,\hspace{0.15em}\theta\hspace{-0.25em}
\in\hspace{-0.25em} [-\varepsilon, \varepsilon],\hspace{0.06em}x\hspace{0.07em})$ and
$f(\theta\,|\,\theta \notin [-\varepsilon, \varepsilon],x)$ that are derived using the type of
methodology outlined in Sections~\ref{sec9} and~\ref{sec10}.
Given the post-data probabilities defined by equations~(\ref{equ12}) and~(\ref{equ13}) and what has
already been assumed, the post-data density of the parameter $\theta$ over the whole of the real
line, which will be denoted as the density $p_0(\theta\,|\,x)$, may now be determined by using the
following expression:
\begin{equation}
\label{equ14}
p_0(\theta\,|\,x) = f(\theta\,|\,\theta \in [-\varepsilon, \varepsilon],x)P_0(\theta \in
[-\varepsilon, \varepsilon]\,|\,x) + f(\theta\,|\,\theta \notin [-\varepsilon,
\varepsilon],x)P_0(\theta \notin [-\varepsilon, \varepsilon]\,|\,x)
\vspace{0.5ex}
\end{equation}
where $f(\theta\,|\,\theta \in [-\varepsilon, \varepsilon],x)$ and $f(\theta\,|\,\theta \notin
[-\varepsilon, \varepsilon],x)$ are the same conditional fiducial densities of $\theta$ that were
used to derive the expected values of $g(x\,|\,\theta_{A})$ and $g(x\,|\,\theta_{B})$ in
equation~(\ref{equ12}).

In the case where the sampling density $g(x_i\,|\,\theta)$ of each independent data value $x_i$ is
a normal density with mean $\theta$ and known variance $\sigma^2$, the fiducial densities
$f(\theta\,|\,\theta \in [-\varepsilon, \varepsilon],x)$ and $f(\theta\,|\,\theta \notin
[-\varepsilon, \varepsilon],x)$ may, for example, be defined according to equations~(\ref{equ8})
and~(\ref{equ9}) or according to equations~(\ref{equ10}) and~(\ref{equ11}).
If these conditional fiducial densities are defined according to these latter two equations and
$\varepsilon > 0$, then the hyperparameter $\tau$\linebreak that appears in the definition of the
GDP function of the parameter $\theta$ in equation~(\ref{equ4}) will be assumed to be chosen so
that the post-data density function of $\theta$ defined by equation~(\ref{equ14}) is a continuous
function rather than a function with discontinuities at $\theta=-\varepsilon$ and
$\theta=\varepsilon$.
This is a particularly convenient criterion for choosing the value of the hyperparameter $\tau$ as
small variations in the value of $\tau$ usually have a very limited effect on the overall shape of
the fiducial density of $\theta$ that is conditioned on $\theta$ lying in the interval
$[-\varepsilon, \varepsilon]$, i.e.\ the density
$f(\theta\,\hspace{0.05em}|\,\hspace{0.05em}\theta\hspace{-0.05em} \in\hspace{-0.05em}
[-\varepsilon, \varepsilon],x)$, and of course no effect, under the assumptions in question, on the
conditional fiducial density $f(\theta\,|\,\theta \notin [-\varepsilon, \varepsilon],x)$.
For a sufficiently large value for the prior probability that $\theta$ lies in the interval
$[-\varepsilon, \varepsilon]$, i.e.\ for a sufficiently large value of $\lambda$, a value of $\tau$
will always exist such that this criterion of ensuring that the post-data density
$p_0(\theta\,|\,x)$ is continuous may be satisfied. In the case where $\varepsilon=0$, the fiducial
density $f(\theta\,|\,\theta \in [-\varepsilon, \varepsilon],x)$ becomes simply a Dirac delta
function, and therefore, it is unavoidable that the post-data density $p_0(\theta\,|\,x)$ will have
a discontinuity at $\theta=0$.

A slightly different way of defining the GPD functions of the parameter $\theta$ on which the
fiducial densities $f(\theta\,|\,\theta \in [-\varepsilon, \varepsilon],x)$ and
$f(\theta\,|\,\theta \notin [-\varepsilon, \varepsilon],x)$ are based to the way that is proposed
in equations~(\ref{equ4}) and~(\ref{equ5}) is presented in Section~5 of Bowater~(2022b) and applied
in Sections~6 and~8 of this earlier paper.
In fact, these alternative definitions of the two GPD functions of $\theta$ in question represent
the limits of the definitions of this function that are given in equations~(\ref{equ4})
and~(\ref{equ5}) as the hyperparameter $\sigma_0$ is allowed to tend to infinity.
The reader is also referred to Bowater~(2022b) for other examples of the application of methodology
that is related to the methodology that has just been developed in the present paper. Furthermore,
it should be pointed out that the methodology put forward in Bowater~(2022b) is generalised in the
appendix of Bowater~(2024).

To illustrate how the problem of inference under discussion may be addressed by using the
expressions in equations~(\ref{equ12}), (\ref{equ13}) and~(\ref{equ14}), the solid curves in
Figures~\ref{fig1} and~\ref{fig2} are plots of the post-data density of the parameter $\theta$
defined by this latter equation with the fiducial densities $f(\theta\,|\,\theta \in [-\varepsilon,
\varepsilon],x$ and $f(\theta\,|\,\theta \notin [-\varepsilon, \varepsilon],x)$ taken to be as
specified by equations~(\ref{equ10}) and~(\ref{equ11}).
In common with the posterior density functions of $\theta$ represented by the long-dashed curves in
these figures, which were derived using the Bayesian method outlined in Section~\ref{sec5}, the
plots of the post-data density of $\theta$ in question were derived under the assumption that the
constant $\varepsilon$ is equal to 0.2, the prior probability that $\theta$ lies in the interval
$[-0.2,0.2]$, i.e.\ the probability $\lambda$, is equal to 0.4, the observed sample mean $\bar{x}$
is equal to 2.576 and the standard error of $\bar{x}$ is equal to one.
Also, similar to the posterior densities of $\theta$ that are represented by the long-dashed curves
in these figures, the assumptions made to derive the two plots of the post-data density of $\theta$
being referred to differ due to the fact that the plot of this post-data density in
Figure~\ref{fig1} corresponds to the parameters $\theta_0$ and $\sigma_0$ of the GPD functions of
$\theta$ in equations~(\ref{equ4}) and~(\ref{equ5}) being equal to zero and 10, respectively, while
in Figure~\ref{fig2}, the plot of this density function corresponds to $\theta_0$ and $\sigma_0$
being equal to 1.5 and one, respectively.

The scenarios on which Figures~\ref{fig1} and~\ref{fig2} are based represent two extremes of the
scenarios that may typically arise in practice.
One of these scenarios corresponds to there being a large difference between the post-data density
function of $\theta$ that is defined by equation~(\ref{equ14}) and the posterior density function
of $\theta$ derived by the Bayesian method described in Section~\ref{sec5}, i.e.\ the scenario
relating to Figure~\ref{fig1}, as indicated by the large difference between the solid and
long-dashed curves plotted in this figure, and the other scenario corresponds to the difference
between the post-data/posterior density functions in question being quite small, i.e.\ the scenario
relating to Figure~\ref{fig2}.
It is natural to compare the two types of post-data/posterior densities of $\theta$ under
discussion in this way since, in Figures~\ref{fig1} and~\ref{fig2}, both of these functions have
been derived under the same assumptions about the data $x$ and also, superficially at least, the
same assumptions about the values of hyperparameters that express pre-data knowledge about the
parameter $\theta$.
Nevertheless, we need to remember, of course, that in deriving a post-data density of $\theta$ of
the type that is defined by equation~(\ref{equ14}), the hyperparameters $\theta_0$ and $\sigma_0$
control the form of a GPD function of $\theta$, while in deriving a posterior density of $\theta$
using the Bayesian method put forward in Section~\ref{sec5}, these hyperparameters control the form
of a prior density~of~$\theta$.

\vspace{3ex}
\subsection{An example of how an incompatible analogy could be introduced}
\label{sec23}

For the sole purpose of giving an example of where we may attempt to address the reference problem
of inference outlined in Section~\ref{sec4} by using multiple analogies that are incompatible, let
us evaluate the possibility of using the Bayesian analogy to determine the post-data densities
$p(\theta_{A}\,|\,x)$ and $p(\theta_{B}\,|\,x)$ on which the expected values of
$g(x\,|\,\theta_{A})$\linebreak and $g(x\,|\,\theta_{B})$ that appear in equation~(\ref{equ12}) are
based. In particular, we will make an analogy between our pre-data knowledge about the values
$\theta_{A}$ and $\theta_{B}$ and what would be our uncertainty about these values if they were
about to be randomly drawn from a population of possible values for $\theta_{A}$ and a population
of possible values for $\theta_{B}$, respectively, where not only do we know that the values of the
parameter $\theta$ in these populations lie inside and outside of the interval $[-\varepsilon,
\varepsilon]$, respectively, but also the compositions of these two populations are completely
known. As an expression of our pre-data knowledge about the values $\theta_A$ and $\theta_B$, it
will be assumed, in accordance with how the Bayesian analogy was defined in Section~\ref{sec1},
that this analogy, which clearly can be viewed as a pair of analogies, will be maintained active
after the data $x$ have been observed.
Therefore, once this analogy or pair of analogies have been made, Bayes' theorem may be applied to
obtain the posterior densities $p(\theta_{A}\,|\,x)$ and $p(\theta_{B}\,|\,x)$, which may then be
substituted into equations~(\ref{equ16}) and~(\ref{equ17}) to determine the expected values of
$g(x\,|\,\theta_{A})$ and $g(x\,|\,\theta_{B})$.

At least at first sight, it would appear that doing this would thereby allow legitimate post-data
probabilities of the parameter $\theta$ lying in the interval $[-\varepsilon, \varepsilon]$ and of
it not lying\linebreak in this interval to be obtained by using equations~(\ref{equ12})
and~(\ref{equ13}), and also allow a justifiable post-data density of $\theta$ over the whole of the
real line to be obtained by using equation~(\ref{equ14}) under the assumption that the fiducial
densities of $\theta$ in this equation are replaced by the posterior densities
$p(\theta_{A}\,|\,x)$ and $p(\theta_{B}\,|\,x)$ under discussion.
However, making the pair of Bayesian analogies that were just mentioned together with the limited
type of Bayesian analogy described in Section~\ref{sec12}, i.e.\ the analogy that justifies the use
of the expressions in equations~(\ref{equ12}), (\ref{equ13}) and~(\ref{equ14}), is equivalent to
addressing the whole of the problem of inference that is being studied by making a single analogy,
namely the standard Bayesian analogy for this problem, which was discussed in Section~\ref{sec5}.
Moreover, addressing this problem of inference by using the method of inference that was outlined
in Section~\ref{sec5}, which is based on this latter analogy, will not, in general, lead to the
posterior probability of the parameter $\theta$ lying in the interval $[-\varepsilon, \varepsilon]$
being equal or even approximately equal to the value that this posterior probability would take if
it was determined by combining multiple Bayesian analogies in the way that has just been described.

Therefore, according to the criterion that was detailed in Section~\ref{sec11} for deciding that a
set of analogies should be considered to be incompatible among themselves in their application to
any given problem of inference, the multiple Bayesian analogies that underlie the method of
inference that was just put forward are incompatible among themselves.
This means of course that the method of inference in question lacks legitimacy and justifiably may
be excluded from further consideration.

Under the assumption that Bayesian inference can only be justified through the use of some form of
the Bayesian analogy as this analogy was defined in Section~\ref{sec1}, the method of inference put
forward in Aitken~(1991), which yields what are called `posterior Bayes factors', is arguably based
on a set of analogies that are essentially a straightforward generalisation of the multiple
Bayesian analogies on which the method of inference just described is based, and is therefore based
on a set of analogies that are incompatible. This is worthy of note since, although the method of
inference proposed in Aitken~(1991) has always been considered to be controversial, it would appear
that it has never before been criticised in this kind of way.
Other criticisms of this method proposed by Aitken can be found in the published comments that
immediately follow Aitken~(1991) and in Bowater~(2022b).

\vspace{3ex}
\subsection{Revisiting Bartlett's paradox}
\label{sec17}

Bartlett's paradox was discussed in Section~\ref{sec13}. Under the same assumptions that were made
in Section~\ref{sec13}, but using more general terminology than was used in this earlier section,
we may describe Bartlett's paradox as being the observation that, independent of what data set $x$
is observed, the post-data probability that the parameter $\theta$ lies in the interval
$[-\varepsilon, \varepsilon]$ tends to one as pre-data belief about where on the real line this
parameter lies becomes less and less concentrated around the value zero, while still maintaining
the assumption that the prior probability of $\theta$ lying in the interval $[-\varepsilon,
\varepsilon]$ is equal to a fixed positive value $\lambda$.
It was pointed out in Section~\ref{sec13} that the Bayesian solution that was proposed in
Section~\ref{sec5} to the reference problem of inference that was outlined in Section~\ref{sec4}
suffers from Bartlett's paradox, which is indeed the case for all standard Bayesian solutions to
this problem of inference.
Let us now present some results to illustrate the fact that the method of inference that is based
on combining different analogies in the way that was put forward in Sections~\ref{sec14}
to~\ref{sec15} does not suffer from this paradox.

In this regard, rows~4 to~6 of Tables~\ref{tab1} to~\ref{tab3} show post-data probabilities of the
parameter $\theta$ lying in the interval $[-\varepsilon, \varepsilon]$ that result from
implementing this method of inference, which is labelled the `fiducial-Bayes' method, in various
scenarios that are relevant to an analysis of Bartlett's paradox.
To be more precise, these post-data probabilities were determined using equation~(\ref{equ12}) with
the conditional fiducial densities $f(\theta\,|\,\theta \in [-\varepsilon, \varepsilon],x)$ and
$f(\theta\,\hspace{0.1em}|\,\hspace{0.1em}\theta\hspace{-0.1em} \notin\hspace{-0.1em}
[-\varepsilon, \varepsilon],x)$ taken to be as specified by equations~(\ref{equ10})
and~(\ref{equ11}).
As was assumed in deriving the results presented in rows~1 to 3 of these tables, the results being
referred to have been calculated under the assumption that the prior probability that $\theta$ lies
in the interval $[-\varepsilon, \varepsilon]$, i.e.\ the probability $\lambda$, is equal to 0.4,
and the standard error of the sample mean, i.e.\ $\sigma/\sqrt{n}$, is equal to one.
Also, the results presented in rows~4 to~6 of Tables~\ref{tab1} to~\ref{tab3} correspond to the
hyperparameter $\theta_0$ that appears in the definitions of the GPD function of the parameter
$\theta$ given in equations~(\ref{equ4}) and~(\ref{equ5}) being equal to zero, and to clarify, the
results under discussion correspond to the variable $\sigma_0$ being the hyperparameter $\sigma_0$
that appears in the definitions of the GPD function of $\theta$ in question.
Finally, let us recall from what was pointed out in Section~\ref{sec13} that the difference in what
was assumed to derive the results presented in Tables~\ref{tab1} to~\ref{tab3} is that these
results correspond to the sample mean $\bar{x}$ being equal to 2.576, 0.8326 and zero,
respectively.

Under the assumptions that have just been made, it can be seen from the rows in Tables~\ref{tab1}
to~\ref{tab3} that correspond to the use of the fiducial-Bayes method that the post-data
probability of $\theta$ lying in the interval $[-\varepsilon, \varepsilon]$ that is defined by
equation~(\ref{equ12}) does not change very much as the hyperparameter $\sigma_0$ increases from
one to infinity for the three different values of $\varepsilon$, namely 0, 0.1 and 0.2, and for the
already mentioned three different values of $\bar{x}$ that are considered in these tables. Also, we
may observe that each of the values of this post-data probability in the rows being referred to in
Tables~\ref{tab1} to~\ref{tab3} is less than the corresponding posterior probability in rows~1 to~3
of these tables that was calculated using the `pure Bayesian' method on the basis of the same
values for the sample mean $\bar{x}$, the constant $\varepsilon$ and the hyperparameter $\sigma_0$.
To justify that this type of comparison of the results of using the fiducial-Bayes method and the
pure Bayesian method is meaningful, it may be pointed out that, for any given value of the
hyperparameter $\sigma_0$, the conditional fiducial density
$f(\theta\,|\,\theta \notin [-\varepsilon, \varepsilon],x)$ that is defined by
equation~(\ref{equ11}) is equal, under the assumptions on which the results in Tables~\ref{tab1}
to~\ref{tab3} are based, to what we obtain by conditioning the posterior density of $\theta$
determined in the Bayesian way via equation~(\ref{equ2}) on $\theta$ lying outside of the interval
$[-\varepsilon, \varepsilon]$.

Furthermore, it can be seen from rows~4 to~6 of Tables~\ref{tab1} to~\ref{tab3} that, when using
the method of inference put forward in Sections~\ref{sec14} to~\ref{sec15}, the limiting value of
the post-data probability of $\theta$ lying in the interval $[-\varepsilon, \varepsilon]$ as the
hyperparameter $\sigma_0$ tends to infinity is far less than one for all values of $\varepsilon$
and all values of $\bar{x}$ that are considered in these tables.
In particular, under the assumption that $\varepsilon=0$, this limiting value of the post-data
probability in question, i.e.\ the post-data probability that $\theta$ equals zero, is 0.0330 when
the sample mean $\bar{x}$ is equal to 2.576, i.e.\ more than twelve times less than the prior
probability that was placed on this hypothesis. This seems reasonable given that, under the null
hypothesis that $\theta$ equals zero, the height of the sampling density of $\bar{x}$ when
$\bar{x}$ equals 2.576 is only 3.62\% of its maximum height.

It is also the case when using the fiducial-Bayes method in question under the assumption of
$\varepsilon$ being equal to zero that the limiting value of the post-data probability of $\theta$
lying in the interval $[-\varepsilon, \varepsilon]$, i.e.\ of $\theta$ being equal to zero, as
$\sigma_0$ tends to infinity is equal to 0.4 when the sample mean $\bar{x}$ takes the value of
0.8326, i.e.\ the same as the prior probability that was placed on this hypothesis, and is equal to
0.4853 when $\bar{x}$ takes the value of zero, i.e.\ greater than the prior probability that was
placed on this hypothesis, which therefore indicates that the observed value of $\bar{x}$
corroborates, to some extent, the hypothesis that $\theta$ equals zero.
These inferences made about the parameter $\theta$ again seem reasonable since, under the null
hypothesis that $\theta$ equals zero, the expected height of the sampling density of $\bar{x}$ at a
value that is randomly drawn from this density function is equal to the height of this sampling
density when $\bar{x}$ is equal to 0.8326, which reveals the reason why this specific value of
$\bar{x}$ was chosen, and the height of this sampling density of $\bar{x}$ when $\bar{x}$ equals
zero is 41.4\% larger than its expected height as calculated in the way just described.

Finally, it is evident from the rows in Tables~\ref{tab1} to~\ref{tab3} that correspond to the use
of the fiducial-Bayes method that, under the assumptions that have been made and with regard to
values of $\sigma_0$ lying in the range one to infinity, the post-data probability of $\theta$
lying in the interval $[-\varepsilon, \varepsilon]$ that is defined by equation~(\ref{equ12}) is
relatively insensitive to changes in the width of the interval $[-\varepsilon, \varepsilon]$ as
$\varepsilon$ goes from zero to 0.2.
This type of insensitivity is generally convenient from a practical viewpoint as when there is a
substantial prior probability that a narrow interval centred on a given value of a parameter of
interest contains the true value of the parameter, making a precise choice for the width of this
interval will often not be easy.

\vspace{3ex}
\subsection{Revisiting Lindley's paradox}
\label{sec18}

Lindley's paradox was discussed in Section~\ref{sec16}.
In particular, it was pointed out in this earlier section that the observation on which this
paradox is based applies to the Bayesian solution that was proposed in Section~\ref{sec5} to the
reference problem of inference that was outlined in Section~\ref{sec4}, which is actually an
observation that applies to all standard Bayesian solutions to this problem of inference, and an
explanation was given as to why this observation may indeed be considered as being a paradox.
Let us now present some results to illustrate the fact that, although standard Bayesian solutions
to the problem of inference in question suffer from Lindley's paradox, the solution to this problem
based on combining different analogies in the way that was put forward in Sections~\ref{sec14}
to~\ref{sec15} does not suffer from this paradox.

In this regard, rows~4 to~6 of Tables~\ref{tab4} and~\ref{tab5} show post-data probabilities of the
parameter $\theta$ lying in the interval $[-\varepsilon, \varepsilon]$ that result from
implementing this method of inference, and therefore from using equation~(\ref{equ12}), in a number
of scenarios that are relevant to the study of Lindley's paradox, where the conditional fiducial
densities $f(\theta\,|\,\theta \in [-\varepsilon, \varepsilon],x)$ and $f(\theta\,|\,\theta \notin
[-\varepsilon, \varepsilon],x)$ on which this equation is based were again taken to be as specified
by equations~(\ref{equ10}) and~(\ref{equ11}).
In common with the assumptions on which the results presented in rows~1 to~3 of these tables are
based, the results being referred to correspond to the prior probability that $\theta$ lies in the
interval $[-\varepsilon, \varepsilon]$, i.e.\ the probability $\lambda$, being equal to 0.4, the
population standard deviation $\sigma$ being equal to 4 and the sample mean $\bar{x}$ always
falling on the 0.995 quantile of the sampling density of $\bar{x}$ under the null hypothesis that
$\theta$ equals zero.
Also, to maintain a degree of comparability with the other results presented in Tables~\ref{tab4}
and~\ref{tab5}, the results presented in rows~4 to~6 of Table~\ref{tab4}, have been calculated by
supposing that the hyperparameters $\theta_0$ and $\sigma_0$ that appear in the definitions of the
GPD function of the parameter $\theta$ given in equations~(\ref{equ4}) and~(\ref{equ5}) are equal
to 0 and 4, respectively, while the results presented in the same rows of Table~\ref{tab5}, have
been calculated by supposing that these hyperparameters are equal to 1.5 and one, respectively.

Under these assumptions, it can be seen from the rows in Tables~\ref{tab4} and~\ref{tab5} that
correspond to the use of the fiducial-Bayes method that, if $\varepsilon$ is equal to zero, the
post-data probability defined by equation~(\ref{equ12}) of the parameter $\theta$ lying in the
interval $[-\varepsilon, \varepsilon]$ i.e.\ of $\theta$ being equal to zero, does not change very
much as the sample size $n$ goes from 10 to infinity for both ways in which the hyperparameters
$\theta_0$ and $\sigma_0$ were specified to derive the results in these tables, and also the
limiting value of this post-data probability as $n$ tends to infinity is equal to 0.0330 in both of
the cases in question.
This limiting value of the post-data probability being referred to is the same as the limiting
value of this probability in the case that is considered in row~4 of Table~\ref{tab1} as the
hyperparameter $\sigma_0$ tends to infinity.

Observe that for the cases where $\varepsilon$ is equal to 0.05 and 0.1 considered in rows~5 and~6
of Tables~\ref{tab4} and~\ref{tab5}, the post-data probability of $\theta$ lying in the interval
$[-\varepsilon, \varepsilon]$ that is defined by equation~(\ref{equ12}) increases substantially as
the sample size $n$ goes from 200 to infinity, and indeed, the limiting value of this post-data
probability as $n$ tends to infinity is one in all of these cases.
However, this is not due to the effects of Lindley's paradox, but due to the fact that by
increasing $n$ when it is already above the value of 200 in the cases in question, the sampling
mean $\bar{x}$ will start getting close to the values in the upper half of the interval
$[-\varepsilon, \varepsilon]$, and actually, for values of $n$ above 43,000, the sampling mean
$\bar{x}$ will lie inside this interval even in the case where $\varepsilon$ is equal to 0.05.

\vspace{3ex}
\section{Combining analogies to weaken Bayesian prior knowledge}
\label{sec20}

\vspace{1.5ex}
\subsection{Methodological description and an illustration}

In many practical situations, it is reasonable to expect that the method of inference that is based
on combining different types of analogies in the way that was put forward in Sections~\ref{sec14}
to~\ref{sec15} will be a useful method for addressing the reference problem of inference that was
described in Section~\ref{sec4}. This is due to the fact that, after pre-data knowledge about the
parameter of interest $\theta$ has been expressed in a way that will often be particularly
convenient, this method of inference can appropriately and efficiently take into account this
pre-data knowledge and the information contained in the observed data in order to make post-data
inferences about the parameter $\theta$.
Also, as was discussed in the previous two sections, this method of inference, unlike the Bayesian
solution that was proposed in Section~\ref{sec5} to the problem of inference in question, does not
suffer from the effects of Bartlett's paradox or Lindley's paradox.

Nevertheless, it could be argued that in a situation where we could, to some meaningful degree,
represent our pre-data knowledge about the parameter $\theta$ by placing a probability distribution
over all possible values for this parameter, the method of inference under discussion may sometimes
substantially overcorrect for the effects of both Bartlett's paradox and Lindley's paradox.
For example, if before the data $x$ were observed, we would have believed that there was some vague
but non-negligible probability that the parameter $\theta$ could lie quite a large distance away
from zero, then in the case where the sample mean $\bar{x}$ happens to take a value that is not
that far from zero, but nevertheless not close to lying in the interval $[-\varepsilon,
\varepsilon]$, we may well feel that there should be a mechanism that increases the post-data
probability of $\theta$ lying in the interval $[-\varepsilon, \varepsilon]$ simply because the
possibility of $\theta$ lying far from zero has been effectively ruled out by the data.
However, due to the fact that a pre-data belief about the parameter $\theta$ of the specific nature
being referred to can not be taken into account by the method of inference outlined in
Sections~\ref{sec14} to~\ref{sec15}, a mechanism for increasing the post-data probability that
$\theta$ lies in the interval $[-\varepsilon, \varepsilon]$ for the reason just highlighted does
not form part of this method of inference.

Motivated by this observation, let us make use of the parameter sampling analogy that was labelled
as Example~\ref{exa2} in Section~\ref{sec6} to construct an alternative way of addressing the
reference problem of inference under discussion.
To clarify, in this analogy, the composition is known of a subset of the population from which the
true value of the parameter $\theta$ is to be randomly drawn, which we will denote as the subset
$K$, while the composition of the subset of this population that contains all the values of
$\theta$ that lie outside of the subset $K$ is either partially or completely unknown. Let this
latter subset be denoted as the subset $U$, and let the proportion that the subset $K$
constitutes of the population in question be denoted as $\kappa$.

In addition to the assumptions on which the parameter sampling analogy under consideration is
based, we will more specifically assume that what we know about the composition of the subset $U$
of the population being referred to before the true value of the parameter $\theta$ is drawn from
this population is equivalent to what we assume that we knew about this parameter before the data
$x$ were observed when using the method of inference put forward in Sections~\ref{sec14}
to~\ref{sec15}.
For example, under the assumption that the conditional fiducial densities
$f(\theta\,\hspace{0.05em}|\,\hspace{0.05em}\theta\hspace{-0.05em} \in\hspace{-0.05em}
[-\varepsilon, \varepsilon],x)$ and $f(\theta\,\hspace{0.05em}|\,\hspace{0.05em}
\theta\hspace{-0.05em} \notin\hspace{-0.05em} [-\varepsilon, \varepsilon],x)$ are as defined in
equations~(\ref{equ8}) and~(\ref{equ9}), what needs to have been known about the parameter $\theta$
before the data $x$ were observed in order for an application of the method of inference in
question to be considered ideal is equivalent to our uncertainty about the parameter $\theta$ if
the true value of this parameter was about to be randomly drawn out of the subset $U$ of the
population just referred to under the assumption that, except for the fact that we know that a
proportion $\lambda$ of the values of $\theta$ in this subset lie in the interval $[-\varepsilon,
\varepsilon]$, the composition of this subset is completely unknown.
Furthermore, in using the parameter sampling analogy being discussed, let us suppose that the
proportion of values in the subset $K$ of the population of values of $\theta$ to be sampled that
lie in the interval $[-\varepsilon, \varepsilon]$ is, as is the case for the subset $U$ of this
population, equal to $\lambda$.

Under these assumptions, if the analogy in question is maintained active after the data $x$ are
observed as an expression of our pre-data knowledge about the parameter $\theta$, then we are
justified in defining the post-data density of $\theta$ as follows:
\begin{equation}
\label{equ15}
p_1(\theta\,|\,x) = \kappa\hspace{0.1em} \pi(\theta\,|\,x) +
(1 - \kappa)\hspace{0.1em}p_0(\theta\,|\,x)
\end{equation}
where $\pi(\theta\,|\,x)$ is the posterior density of $\theta$ that is determined by using the
standard Bayesian method of inference outlined in Section~\ref{sec5}, i.e.\ determined according to
equation~(\ref{equ2}), and $p_0(\theta\,|\,x)$ is the post-data density of $\theta$ that is
determined by using the meth-{\linebreak}od of inference outlined in Sections~\ref{sec14}
to~\ref{sec15}, i.e.\ determined according to equation~(\ref{equ14}). Therefore, the post-data
density $p_1(\theta\,|\,x)$ defined by the equation just presented is simply a mixture of the
post-data densities $\pi(\theta\,|\,x)$ and $p_0(\theta\,|\,x)$.

In the context of the analogy being made, the smaller the value of $\kappa$, the smaller is the
subset $K$ relative to the subset $U$ of the population from which the true value of the parameter
$\theta$ is to be drawn, which implies the lower is the level of concrete knowledge that we have
about the composition of this population, and as a result, the more the post-data density of
$\theta$ should resemble the post-data density $p_0(\theta\,|\,x)$ defined by
equation~(\ref{equ14}) rather than the posterior density $\pi(\theta\,|\,x)$ derived using the
standard Bayesian method.
Therefore, with its maximum value of one as the starting point, progressively reducing the value of
$\kappa$ can be viewed as a process of weakening the strength of the knowledge that we had about
the parameter $\theta$ before the data $x$ were observed that was of a Bayesian nature.
This is arguably a more fundamental way of weakening the strength of this type of pre-data
knowledge than the common strategy of using the standard Bayesian method of inference, but making
the prior distribution of the parameter of interest more diffuse over the real line. The reason for
this is that, in the analogy that naturally needs to be made to justify this latter strategy, it
still has to be assumed that, before the data are observed, the population from which the true
value of the parameter of interest is randomly drawn has a composition that is completely known.

Similar to how the choices need to made for the values of the other hyperparameters that control
the expression of pre-data knowledge in constructing the post-data density $p_1(\theta\,|\,x)$
given by equation~(\ref{equ15}), for example, the value of $\lambda$ and the values of the
hyperparameters that appear in equations~(\ref{equ3}), (\ref{equ4}) and~(\ref{equ5}), the value of
$\kappa$ needs to be chosen so that by means of the parameter sampling analogy just described we
may best represent what we genuinely knew about the parameter $\theta$ before the data $x$ were
observed.
Of course, this representation of our pre-data knowledge about the parameter $\theta$ will usually
be, to some degree, imperfect, but the same comment could be made about all the other analogies for
representing pre-data knowledge about a parameter of interest that have been considered so far
in the present paper.
Also, it is worth pointing out that if the GPD functions of the parameter $\theta$ on which the
post-data density $p_0(\theta\,|\,x)$ is based are chosen as carefully as possible to independently
represent the pre-data knowledge that there was about this parameter, then in practice, if a value
of zero is not an appropriate choice for $\kappa$, then it could be anticipated that choosing
$\kappa$ to be in the range of 0 to 0.3 is likely to be reasonable in most situations with it being
unlikely that situations will often arise where $\kappa$ would need to be greater than 0.5.

To illustrate the use of the method of inference that was just put forward, let us begin by
assuming that the sampling density $g(x_i\,|\,\theta)$ of each independent data value $x_i$ is a
normal density with mean $\theta$ and known variance $\sigma^2$, and the prior density of $\theta$
on which the posterior density $\pi(\theta\,|\,x)$ is based is as defined in equation~(\ref{equ3}).
In addition, we will suppose that the conditional fiducial densities $f(\theta\,|\,\theta \in
[-\varepsilon, \varepsilon],x)$ and $f(\theta\,|\,\theta \notin [-\varepsilon, \varepsilon],x)$
that underlie the construction of the post-data density $p_0(\theta\,|\,x)$ are as defined in
equations~(\ref{equ10}) and~(\ref{equ11}).
Under these assumptions, the dotted curve in Figure~\ref{fig1} is a plot of the post-data density
of $\theta$ defined by equation~(\ref{equ15}).
In common with the post-data densities of $\theta$ represented by the long-dashed and solid curves
in this figure, the plot of the post-data density of $\theta$ that is of current interest was
derived by taking the constant $\varepsilon$ to be 0.2, the prior probability $\lambda$ to be 0.4,
the observed sample mean $\bar{x}$ to be 2.576 and the standard error of $\bar{x}$ to be one.
Also, in deriving the plot in question, it was supposed that the values of the parameters
$\theta_0$ and $\sigma_0$ of the GPD functions of $\theta$ defined by equations~(\ref{equ4})
and~(\ref{equ5}) take the same values as the parameters $\theta_0$ and $\sigma_0$ of the prior
density of $\theta$ defined by equation~(\ref{equ3}), and these values, similar to what was assumed
in deriving the other two post-data densities of $\theta$ that are plotted in Figure~\ref{fig1},
were assumed to be equal to zero and 10, respectively.
Finally, the constant $\kappa$ that modifies the parameter sampling analogy that was just discussed
was assumed to be equal to 0.2.

It should be clear that, as a result of the assumptions made to construct the plots of the
post-data densities of $\theta$ that are presented in Figure~\ref{fig1}, the long-dashed and solid
curves in this figure are, respectively, the post-data densities $\pi(\theta\,|\,x)$ and
$p_0(\theta\,|\,x)$ that define, by means of equation~(\ref{equ15}), the post-data density
$p_1(\theta\,|\,x)$ that is represented by the dotted curve in this figure, which implies, of
course, that this latter post-data density of $\theta$ is simply a mixture of the post-data
densities of $\theta$ that are represented by the long-dashed and solid curves in question.

\vspace{3ex}
\subsection{Revisiting Bartlett's and Lindley's paradoxes again}

As was the intention when designing the method of inference that was outlined in the previous
section, it can be argued that, in certain situations, this method of inference avoids the effects
of Bartlett's and Lindley's paradox in a way that is better than the method of inference that was
outlined in Sections~\ref{sec14} to~\ref{sec15}.
To illustrate this point, the final rows of Tables~\ref{tab1} to~\ref{tab5} show, for the case
where $\varepsilon$ is equal to zero, post-data probabilities of the parameter $\theta$ lying in
the interval $[-\varepsilon, \varepsilon]$, i.e.\ being equal to zero, that result from
implementing the method of inference put forward in the preceding section, which is labelled the
`mixture' method, in various different scenarios.
These results have been derived under the same assumptions that were used to derive the
corresponding results in each of the tables in question for what are referred to, in these tables,
as the `pure Bayesian' and `fiducial-Bayes' methods, i.e.\ the assumptions outlined in
Sections~\ref{sec13} and~\ref{sec17} for Tables~\ref{tab1} to~\ref{tab3} and the assumptions
outlined in Sections~\ref{sec16} and~\ref{sec18} for Tables~\ref{tab4} and~\ref{tab5}.
The only extra assumption that needed to be made to derive the post-data probabilities being
referred to was to assume that the parameter $\kappa$ that modifies the parameter sampling analogy
on which the method of inference proposed in the last section is based takes the value of 0.2 in
every case of interest. These assumptions imply that each result reported for the mixture method in
the final rows of Tables~\ref{tab1} to~\ref{tab5} is simply a weighted average of the results for
the pure Bayesian and fiducial-Bayes methods that correspond to $\varepsilon$ being equal to zero
and that appear in the corresponding column of the table concerned, with the weight on the relevant
result for the fiducial-Bayes method being four times larger than the relevant result for the
pure-Bayesian method.

On the basis of the post-data probabilities that are reported in the final rows of
Tables~\ref{tab1} to~\ref{tab5}, it may be appreciated that the method of inference put forward in
the previous section may represent a satisfactory compromise between the fiducial-Bayes method of
inference outlined in Sections~\ref{sec14} to~\ref{sec15} and the pure Bayesian method of inference
outlined in Section~\ref{sec5} in situations where pre-data knowledge about the parameter $\theta$
can not be adequately expressed simply in terms of a prior probability $\lambda$ of the parameter
$\theta$ lying in the interval $[-\varepsilon, \varepsilon]$ and two GPD functions of $\theta$, one
of which is conditioned on $\theta$ lying in the interval $[-\varepsilon, \varepsilon]$ and the
other is conditioned on $\theta$ lying outside of this interval, e.g.\ the GPD functions of
$\theta$ given in equations~(\ref{equ4}) and~(\ref{equ5}).

\vspace{3ex}
\section{Conclusion and further discussion}
\label{sec21}

This paper has made the case that analogy making should be the foundation on which methods of
statistical inference are constructed and the principal means by which they are justified, which in
effect implies that analogy making offers a way of resolving what, at the start of this paper, was
defined as being the fundamental problem of statistical inference.
More specifically, it has been argued that preferences among competing methods of inference should
be determined primarily on the basis of how appropriate it is to make the analogies on which these
methods are based in the context of the problem of inference that is trying to be addressed.
Also, it may often be convenient to use different analogies to address different parts of a problem
of inference, provided that it has, of course, been checked that the analogies are compatible among
themselves. Finally, the case may be made that methods of inference should be classified as being
distinct from each other if the analogies on which they are based are not equivalent.
Given that this latter point is important and has not been discussed to a great extent so far, let
us conclude this paper by exploring it in a little more detail.

To begin with, let us consider using the Bayesian analogy to make inferences about the mean
$\theta$ of a normal distribution that has a known variance $\sigma^2$ on the basis of a sample $x$
drawn from the distribution concerned. In doing this, if it is assumed that the prior distribution
of $\theta$ is a normal distribution with a mean of $\theta_0$ and a variance of $\sigma_0^2$, then
the posterior distribution of $\theta$ would be a normal distribution with a mean of $\theta_1$ and
a variance of $\sigma_1^2$, where $\theta_1$ and $\sigma_1^2$ are as defined in
equations~(\ref{equ18}) and~(\ref{equ19}) with the hyperparameters $\theta_0$ and $\sigma_0^2$ in
these equations obviously now being parameters of a prior distribution of $\theta$ rather than
parameters of the GPD functions of $\theta$ in equations~(\ref{equ4}) and~(\ref{equ5}). Having made
this assumption, let us now point out the well-known fact that if, while keeping all other
variables fixed, $\sigma_0^2$ is allowed to tend to infinity, then the posterior distribution of
$\theta$ just referred to will tend to the fiducial distribution of $\theta$ that is defined by
equation~(\ref{equ20}).

On the basis of this observation, it is often claimed that using the standard fiducial method of
inference to address the problem of inference in question is equivalent, or approximately
equivalent, to addressing this problem using a Bayesian method of inference, see for example, Liu
and Martin~(2015). However, it is clear that the Bayesian analogy that is required to derive the
type of approximation being highlighted to the post-data density of $\theta$ in
equation~(\ref{equ20}) is completely different from the fiducial analogy that was used to derive
this post-data density of $\theta$ in Section~\ref{sec2}.
Furthermore, while making this fiducial analogy is likely to be considered as being very acceptable
in many situations, it would appear that much fewer situations are likely to arise in practice
where one would feel reasonably comfortable in making the type of Bayesian analogy just mentioned.
Therefore, it has hopefully been put beyond any doubt that using the fiducial method of inference
that was described in Section~\ref{sec2} to address the problem of inference under discussion is
not equivalent, or approximately equivalent, to addressing this problem by using a Bayesian method
of inference.

Let us now turn our attention to addressing the same problem of inference using an analogy that we
will choose to call Fisher's analogy due to it being possible to view this analogy as underlying
the way R.\ A.\ Fisher attempted to justify fiducial inference from a frequentist perspective in
his final book on statistical inference, namely Fisher~(1956). We will begin by defining this
analogy.

\vspace{3ex}
\noindent
\textbf{Fisher's analogy}

\vspace{1.5ex}
\noindent
To introduce this analogy, let us assume that by using any data set $x$ that is randomly generated
from a given sampling distribution of interest as an input to a given set of calculations, an
interval $I_x$ of the real line may be determined. Also, we will suppose that a known proportion
$\beta$ of the intervals $I_x$ that are generated in this way will contain the true value of a
given parameter $\theta$ of the sampling distribution concerned, no matter what are the true values
of this parameter and other unknown parameters on which this sampling distribution may depend.

Under these assumptions, Fisher's analogy is an analogy made after a given data set $x$ has been
observed between the proportion $\beta$ and the probability that the interval $I_x$ calculated on
the basis of the observed data set $x$ will contain the true value of the parameter $\theta$.
In particular, it is based on the argument that if the information contained in the observed data
$x$, or pre-data knowledge about the parameter $\theta$ and other unknown parameters of the
sampling distribution of interest, does not make it more or less likely that the observed interval
$I_x$ will contain the true value of $\theta$ than any other interval $I_x$ that could have been
observed, then the probability that the true value of $\theta$ lies in the observed interval $I_x$
should be judged as being equal to the proportion $\beta$.

\vspace{3ex}
For example, let us imagine that, in addressing the problem of inference that was just being
discussed, we have calculated, for a given value of $\beta$, the standard
$100\hspace{0.05em}\beta\hspace{0.05em}\%$ confidence interval of the parameter $\theta$ for this
problem, i.e.\ the confidence interval:
\begin{equation}
\label{equ22}
\left( \bar{x} - \Phi^{-1} ((1+\beta)/2)(\sigma/\sqrt{n}\hspace{0.1em}),\hspace{0.2em}
\bar{x} + \Phi^{-1} ((1+\beta)/2)(\sigma/\sqrt{n}\hspace{0.1em}) \right)
\end{equation}
on the basis of the observed data set $x$.
In this scenario, if nothing or very little was known about the parameter $\theta$ before the data
set $x$ was observed, then after considering all other possible data sets $x$ that could have been
observed and, for the same value of $\beta$, all confidence intervals of the type in question that
would have been calculated on the basis of each of these data sets, we may well decide that making
Fisher's analogy is appropriate in this case, and thereby judge that the post-data probability of
$\theta$ lying in the observed confidence interval defined by equation~(\ref{equ22}) should be
equal to the proportion~$\beta$.

Given that integrating the post-data density of $\theta$ in equation~(\ref{equ20}) over the
confidence interval in equation~(\ref{equ22}) yields a post-data probability of the parameter
$\theta$ lying in this interval that is equal to $\beta$, it can be seen that using the fiducial
analogy described in Section~\ref{sec2} to make post-data inferences of the type of current
interest about the parameter $\theta$ yields the same numerical result as is obtained by using what
is being called Fisher's analogy.
Nevertheless, it is clear that the fiducial analogy is very different from Fisher's analogy.
For example, in the former type of analogy, the random variable of main interest is a variable that
conforms to the definition of a primary random variable given in Bowater~(2019, 2021), while in the
latter type of analogy, it is the data $x$ themselves that constitute the random variable of main
interest.
Also, in deciding whether the fiducial analogy is an analogy that is adequate to make, we only need
to consider the data set $x$ that was actually observed, while in deciding whether it is adequate
to make Fisher's analogy, we also need to consider all the data sets $x$ that could have been
observed.
In fact, Fisher's analogy is not even a pre-post event analogy according to how this general type
of analogy was defined in Section~\ref{sec2}, since what is in effect the event of main interest
before the data $x$ are observed, namely the event of the known true value of the parameter
$\theta$ lying in a randomly generated interval $I_x$, is not the same as the event of main
interest after the data $x$ are observed, namely the event of the unknown true value of $\theta$
lying in a given fixed interval $I_x$.

Furthermore, while it is not that difficult, as may be appreciated from Bowater~(2018, 2019, 2021,
2022b, 2023), to use the fiducial analogy to address problems of inference that are much more
complicated that the one that was just discussed, this is not the case for Fisher's analogy.
To clarify this point, let us begin by noting that in trying to apply Fisher's analogy to any given
problem of inference, it may be possible to establish that the observed data set $x$ belongs to
various subsets of the set of all data sets that could have been observed, which we will denote as
the subsets $\mathcal{R}_1, \mathcal{R}_2, \ldots$, with the property that if randomly generated
data sets $x$ are conditioned to lie in any one of these subsets, then the proportion of intervals
$I_x$ calculated on the basis of these data sets that contain the true value of the parameter
$\theta$ will either be strictly greater or strictly less than $\beta$ over all possible values of
$\theta$ and other unknown parameters on which the sampling distribution of interest may depend.
Therefore, why should we assign a post-data probability of $\beta$ to the event of the parameter
$\theta$ lying in the interval $I_x$ calculated on the basis of the observed data set $x$ when we
are faced with the fact that, if we chose to regard this data set as being a typical member of any
one of the subsets $\mathcal{R}_1, \mathcal{R}_2, \ldots$, then the proportion of intervals $I_x$
that would contain the true value of $\theta$ under the requirement that such intervals are
calculated using randomly generated data sets $x$ conditioned to lie in the chosen subset would be
strictly greater or strictly less than $\beta$ over all possible states of the world?
Clearly, the existence of a misgiving of this nature is likely to lead to the application of
Fisher's analogy to be considered invalid.

In fact, in the terminology of Fisher~(1956), the subsets $\mathcal{R}_1, \mathcal{R}_2, \ldots$
just referred to would be called \textit{recognisable subsets} of the sample space, and generally
it is agreed that being able to identify subsets of the sample space of this type is usually
sufficient to regard the application of the version of fiducial inference that was proposed in
Fisher~(1956) as being invalid (see, for example, Robinson 1979), and therefore, in effect, the use
of Fisher's analogy as it is being defined here as being invalid.
Moreover, it should be pointed out that we do not need to move very far away from the simplest
problems of statistical inference, before we encounter problems where the application of Fisher's
analogy is hampered by the existence of recognisable subsets of the sample space.
For example, as shown in Buehler and Feddersen~(1963) and Brown~(1967), it is possible to identify
recognisable subsets of the sample space even when trying to use the ideas in Fisher~(1956) to give
post-data validity to the standard confidence interval of the mean $\mu$ of a normal distribution
in the case where its variance $\sigma^2$ is also an unknown parameter, i.e.\ the standard Student
$t$ confidence interval of a normal mean $\mu$.

By means of the discussion just presented, it is hoped that it has been adequately demonstrated
that the simple fact that the type of fiducial inference put forward in Bowater~(2019, 2021), i.e.\
organic fiducial inference, is based on the fiducial analogy rather than Fisher's analogy is
sufficient for it to be regarded as being completely distinct from the version of fiducial
inference proposed in Fisher~(1956), and more generally, going back to the original motivation for
this discussion, that any two methods of inference may be regarded as being distinct from each
other simply due to the fact that they are based on analogies that are not equivalent.

\vspace{5.5ex}
\pdfbookmark[0]{References}{toc1}
\noindent
\textbf{References}

\begin{description}

\setlength{\itemsep}{1ex}

\vspace{0.5ex}
\item[] Aitken, M. (1991).\ Posterior Bayes factors (with discussion).\ \textit{Journal of the
Royal Statistical Society, Series B}, \textbf{53}, 111--142.

\item[] Bartlett, M. S. (1957).\ A comment on D. V. Lindley's statistical paradox.\
\textit{Biometrika}, \textbf{44}, 533--534.

\item[] Bernardo, J. M. and Smith, A. F. M. (1994).\ \textit{Bayesian Theory}, Wiley, New York.

\item[] Bowater, R. J. (2018).\ Multivariate subjective fiducial inference.\ \textit{arXiv.org
(Cornell University), Statistics}, arXiv:1804.09804.

\item[] Bowater, R. J. (2019).\ Organic fiducial inference.\ \textit{arXiv.org (Cornell
University), Sta\-tis\-tics}, arXiv:1901.08589.

\item[] Bowater, R. J. (2021).\ A revision to the theory of organic fiducial inference.\
\textit{arXiv.org (Cornell University), Statistics}, arXiv:2111.09279.

\item[] Bowater, R. J. (2022a).\ Physical, subjective and analogical probability.\
\textit{arXiv.org (Cornell University), Statistics}, arXiv:2204.10159.

\item[] Bowater, R. J. (2022b).\ Sharp hypotheses and organic fiducial inference.\
\textit{arXiv.org (Cornell University), Statistics}, arXiv:2207.08882.

\item[] Bowater, R. J. (2023).\ The fiducial-Bayes fusion:\ a general theory of statistical
inference.\ \textit{arXiv.org (Cornell University), Statistics}, arXiv:2310.01533.

\item[] Bowater, R. J. (2024).\ Probabilistic inference when the population space is open.\\
\textit{arXiv.org (Cornell University), Statistics}, arXiv:2410.12930.

\item[] Brown, L. (1967).\ The conditional level of Student's t test.\ \textit{Annals of
Mathematical Statistics}, \textbf{38}, 1068--1071.

\item[] Buehler, R. J. and Feddersen, A. P. (1963).\ Note on a conditional property of
Student's~t.\ \textit{Annals of Mathematical Statistics}, \textbf{34}, 1098--1100.

\item[] de Finetti, B. (1937).\ La pr\'evision:\ ses lois logiques, ses sources subjectives.\
\textit{Annales de l'Institut Henri Poincar\'e}, \textbf{7}, 1--68.\ Translated into English in
1964 as `Foresight:\ its logical laws, its subjective sources' in \textit{Studies in Subjective
Probability}, Eds.\ H. E. Kyburg and H. E. Smokler, Wiley, New York, pp.\ 93--158.

\item[] Fisher, R. A. (1956).\ \textit{Statistical Methods and Scientific Inference}, Hafner Press,
New York [2nd ed., 1959; 3rd ed., 1973].

\item[] Jaynes, E. T. (2003).\ \textit{Probability Theory:\ The Logic of Science}, Cambridge
University Press, Cambridge.

\item[] Levi, I. (1985).\ Imprecision and indeterminacy in probability judgment.\
\textit{Philosophy of Science}, \textbf{52}, 390--409.

\item[] Lindley, D. V. (1957).\ A statistical paradox.\ \textit{Biometrika}, \textbf{44}, 187--192.

\item[] Liu, C. and Martin, R. (2015).\ Frameworks for prior-free posterior probabilistic
inference.\ \textit{WIREs Computational Statistics}, \textbf{7}, 77--85.

\item[] Robinson, G. K. (1979).\ Conditional properties of statistical procedures.\ \textit{Annals
of Statistics}, \textbf{7}, 742--755.

\item[] Savage, L. J. (1954).\ \textit{The Foundations of Statistics}, Wiley, New York.

\item[] Shafer, G. (1976).\ \textit{A Mathematical Theory of Evidence}, Princeton University Press,
Princeton.

\item[] Walley, P. (1991).\ \textit{Statistical Reasoning with Imprecise Probabilities}, Chapman
and Hall, London.

\end{description}

\end{document}